\documentclass[11pt,a4paper]{article}

\usepackage[utf8]{inputenc}
\usepackage[T1]{fontenc}
\usepackage{amsmath,amssymb}
\usepackage{graphicx}
\usepackage{booktabs}
\usepackage{float}
\usepackage{caption}
\usepackage{xspace}
\usepackage[margin=2.5cm]{geometry}
\usepackage[numbers,sort&compress]{natbib}
\usepackage[colorlinks=true,linkcolor=blue,citecolor=blue,urlcolor=blue]{hyperref}

\title{Can torsion mimic a cosmological constant? A one-parameter logarithmic $f(T)$ extension of $\Lambda$CDM}

\author{
A. O. Aquino\thanks{Corresponding author: \texttt{adrieloliveira@fisica.ufc.br}} \quad and \quad J.E.G. Silva\thanks{\texttt{euclides@fisica.ufc.br}}\\[0.3cm]
\small Department of Physics, Federal University of Ceará, Fortaleza, Ceará, Brazil
}
\date{}

\begin{document}

\maketitle

\begin{abstract}
We study a modified teleparallel gravity model of the $f(T)$ type. We propose a new functional form for in order to obtain a geometric origin to the $\Lambda CDM$ behavior in an appropriate limit. A perturbative expansion around the  $\Lambda CDM$ regime is carried out, and the model's cosmic dynamics are calculated inside the flat FLRW background. By employing a Markov Chain Monte Carlo (MCMC) analysis using multiple observational datasets, such as Cosmic Chronometers, Type Ia Supernova compilations (DES-SN5YR, Pantheon+SH0ES, and Union 3.0), DESI DR2 BAO measurements, and Planck 2018 CMB distance priors, to constrain the model parameters. The findings show that the suggested $f(T)$ model accurately replicates the observed accelerated expansion and is statistically consistent with current cosmological observations. The model stays close to the  $\Lambda CDM$ scenario while permitting tiny dynamical adjustments, as evidenced by the extra parameter $\gamma$, which quantifies deviations from the conventional cosmological model, being limited to values near zero.
\end{abstract}


\section{Introduction}
\label{sec1}

The concept of dark energy remains one of the major problems in cosmology, as many hypotheses have been proposed, however, none have been definitively confirmed. The cosmological constant $\Lambda$ in the standard cosmological model, $\Lambda$CDM, is still our best theoretical description. This model successfully fits a wide range of observational data, including Type Ia supernova like Pantheon+SH0ES \cite{Scolnic_2022}, Union 3.0 \cite{Rubin_2025} and DES-SN5YR \cite{DES_2024}, the Cosmic Microwave Background (CMB) of Planck Collaboration \cite{Planck_2018_params}, Baryon Acoustic Oscillations (BAO) from DESI DR2 \cite{Adame2025DESI}, and others. However, the cosmological constant faces conceptual and observational challenges. One of the most significant is the vacuum energy problem, which stems from the vast discrepancy between the vacuum energy density predicted by quantum field theory and the extremely small value inferred from cosmological observations---differing by roughly 120 orders of magnitude \cite{RevModPhys.61.1,Padmanabhan_2003}. Furthermore, $\Lambda$ introduces no intrinsic dynamics, being a fixed parameter with an equation of state $w = -1$, offering no explanation for its small but nonzero value or for the apparent coincidence that its effects dominate the cosmic expansion only in the present epoch \cite{Carroll_2001}. In addition, recent observations have revealed tensions within the $\Lambda$CDM framework, notably between early- and late-universe determinations of the Hubble constant is not the same value in different ages of universe, in eraly-universe the CMB give us $H_0 = 67.4 \pm 0.5$ $km$ $s^{-1}Mpc^{-1}$ \cite{Planck_2018_params}, and in late-universe the Pantheon+SH0ES give us $73.04 \pm 1.04$ $km$ $s^{-1}Mpc^{-1}$ \cite{Scolnic_2022}. These theoretical inconsistencies and observational tensions suggest that the cosmological constant may be an oversimplification, motivating the exploration of dynamical models for dark energy or modified gravity theories as possible extensions to the standard cosmology.\\

The modified models have been proposed to explain the cosmic expansion without the cosmological constant. They based on different extensions of Einstein-Hilbert action: functions of curvature scalar $R$ in case of $f(R)$ models \cite{Bessa_2022,Kumar_2025}, functions of torsion scalar $T$ in case of $f(T)$ teleprallel models \cite{Sandoval_2024,Zhadyranova_2024} or funcions of Nonmetricity scalar $Q$ in case of $f(Q)$ models \cite{Myrzakulov_2023,Koussour_2023}. These models produce good fits for a late-time universe in accelerated expansion, however, they typically introduce additional free parameters and lead to more complex Friedmann equations, making the theoretical and numerical analyses more challenging.\\

This paper is organized as follows: In Section~\ref{sec2}, we briefly review the fundamentals of Teleparallel Gravity and its standard cosmological formulation. Section~\ref{sec3} presents the modified $f(T)$ gravity framework, introducing our novel $f(T)$ model motivated by a quasi-constant energy density. In Section~\ref{sec4}, we describe the observational datasets, including Cosmic Chronometers, Type Ia Supernovae, DESI DR2 BAO, and Planck 2018 CMB distance priors along with the MCMC numerical setup. In Section~\ref{sec5}, we present and discuss the observational constraints and compare our $f(T)$ model with the standard $\Lambda$CDM scenario. Finally, Section~\ref{sec6} summarizes our main conclusions and final remarks.

\section{Teleparallel gravity and cosmology}
\label{sec2}
Teleparallel gravitational is one alternative theory of gravitation based on the torsion of spacetime, instead of working with a curvature scalar. For this, we will use the tetrads $e_{a}(x^{\mu})$, where $a=0,1,2,3$ describe vectors that form an orthonormal basis in a tangent space for each point $x^{\mu}$ of a manifold. The vectors $e_{a}$ are orthogonal to each other, so $e_{a} \cdot e_{b} = \eta_{ab}$, where $\eta_{ab}$ is the Minkowski metric of the local tangent space given by $\eta_{ab} = diag(-1,1,1,1)$, that is, for each point $x^{\mu}$ in a curved space, we will have a correspondence in a flat tangent space.\\

Each tetrad $e_{a}$ has components $e_{a}^{\mu}$, where $\mu = 0,1,2,3$ represent coordinates of the curved spacetime, thus we can express the vector as $e_{a}= e^{\mu}_{a}\partial_{\mu}$, thus the Latin indices will be components of the tangent spacetime and the Greek indices will be components of the curved manifold. As in the usual approach of the Theory of General Relativity, we need a metric, we can express it as $g_{\mu\nu}= \eta_{ab} e^{a}_{\mu}e^{b}_{\nu}$, thus we have a direct relationship between the metric of the curved manifold and the flat metric of the tangent spacetime. We will also have that the inverse will obey the relation $e^{a}_{\mu}e^{\mu}_{b} = \delta^{a}_{b}$, where $\delta^{a}_{b}$ is the Kronecker delta. By the metric too, we can define that $\sqrt{-g}= det[e^{a}_{\mu}] = e$. If we are changing the way of describing gravitational effects in space-time, it is clear that there will also be changes in the way of writing the connection $\Gamma$. While in General Relativity it is common to use the symmetric connection known as the Levi-Civita connection, based on the tetrad postulate $\nabla_\nu e^a{}_\mu = 0$, in the covariant formulation \cite{Krssak:2015oua} of teleparallel gravity our connection incorporates the flat spin connection $\omega^a{}_{b\nu}$ and is represented in Eq.(\ref{weitzenbock})
\begin{equation}
\label{weitzenbock}
    \Gamma^{\lambda}{}_{\mu\nu} = e^{\lambda}{}_{a} \left( \partial_{\nu} e^{a}{}_{\mu} + \omega^{a}{}_{b\nu} e^{b}{}_{\mu} \right).
\end{equation}
Accordingly, the torsion tensor is given by
\begin{equation}
\label{torsion}
    T^{\lambda}{}_{\mu\nu} = \Gamma^{\lambda}{}_{\mu\nu} - \Gamma^{\lambda}{}_{\nu\mu} = e^{\lambda}{}_{a} \left( \partial_{\mu} e^{a}{}_{\nu} - \partial_{\nu} e^{a}{}_{\mu} + \omega^{a}{}_{b\mu} e^{b}{}_{\nu} - \omega^{a}{}_{b\nu} e^{b}{}_{\mu} \right).
\end{equation}
From Eq.(\ref{torsion}) we can define the torsion scalar as \cite{Bamba_2012}
\begin{equation}
\label{torsion_scalar}
    T = T^{\lambda}{}_{\mu\nu} S_{\lambda}{}^{\mu\nu},
\end{equation}
where the tensor $S_{\lambda}{}^{\mu\nu}$ is defined as
\begin{equation}
\label{superpotential}
    S_{\lambda}{}^{\mu\nu} = \frac{1}{2} \left( K_{\lambda}{}^{\mu\nu} + \delta^{\mu}_{\lambda} T^{\sigma\nu}{}_{\sigma} - \delta^{\nu}_{\lambda} T^{\sigma\mu}{}_{\sigma} \right).
\end{equation}
And the tensor $K_{\lambda}{}^{\mu\nu}$ is called the contorsion tensor, defined as the difference between the Weitzenböck and Levi-Civita connections, respectively, as represented in Eq.(\ref{contorsion})
\begin{equation}
\label{contorsion}
    K_{\lambda}{}^{\mu\nu} = \Gamma^{\lambda}{}_{\mu\nu} - {}^{\circ}\Gamma^{\lambda}{}_{\mu\nu}.
\end{equation}

It is worth noting that this is the covariant formulation of teleparallelism, in which the spin connection $\omega^a{}_{b\nu}$ is explicitly incorporated into the teleparallel connection. In this way, the tetrad and the spin connection transform jointly under local Lorentz transformations, ensuring the covariance of the formulation and eliminating the frame dependence associated with the pure-tetrad formulation. In the latter, the Weitzenböck gauge is adopted, where the spin connection vanishes ($\omega^a{}_{b\nu}=0$), causing the teleparallel connection to reduce to $\Gamma^\lambda{}_{\mu\nu} = e^\lambda{}_a \partial_\nu e^a{}_\mu$.

In teleparallel gravity we have the action as Eq.(\ref{T_Einstein})
\begin{equation}
\label{T_Einstein}
    S = \frac{1}{2\kappa^2} \int d^4 x \, e (T - 2\Lambda) + \int d^4 x \, e (\mathcal{L}_m + \mathcal{L}_r),
\end{equation}
and if we make the variation of Eq.(\ref{T_Einstein}) with respect to the tetrad vector field we have the field equation for teleparallel gravity in Eq.(\ref{field_equation}) \cite{Krssak:2015oua}
\begin{equation}
\label{field_equation}
    e^{-1}\mathcal{D}_{\mu}\left(e \, S_{a}{}^{\mu\nu}\right)
    + e_{a}{}^{\lambda} T^{\rho}{}_{\mu\lambda} \, S_{\rho}{}^{\nu\mu}
    + \frac{1}{4} \, e_{a}{}^{\nu} T
    + \Lambda \, e_{a}{}^{\nu}
    = \kappa^{2} \mathcal{T}_{a}{}^{\nu},
\end{equation}
where $\mathcal{D}_\mu S_a{}^{\mu\nu} = \partial_\mu S_a{}^{\mu\nu} - \omega^b{}_{a\mu} S_b{}^{\mu\nu}$ is the Lorentz covariant derivative, $\kappa^2 = 8 \pi G$ and $\mathcal{T}_a{}^\nu$ is the stress--energy tensor. If we use the flat Friedmann-Lemaître-Robertson-Walker (FLRW) metric given by Eq.(\ref{flrw})
\begin{equation}
\label{flrw}
    ds^2 = -c^2dt^2 + a^2(t)(dx^2 + dy^2 + dz^2),
\end{equation}
which corresponds to the tetrad choice
\begin{equation}
\label{tetrad}
    e^{a}_\mu = \mathrm{diag}(1, a(t), a(t), a(t)),
\end{equation}
our torsion scalar becomes $T = -6 \left(\frac{\dot{a}}{a} \right)^2$, and solving Eq.(\ref{field_equation}) for FLRW metric we have the Friedmann usual equations
\begin{gather}
\label{friedmann}
\left(\frac{\dot{a}}{a}\right)^2 = H^2 = \frac{8\pi G}{3}\rho_i + \frac{\Lambda}{3}, \\[0.3cm]
\frac{\ddot{a}}{a} = -\frac{4\pi G}{3}(\rho_i + 3p_i) + \frac{\Lambda}{3},
\end{gather}
like usual General Relativity with curvature.

\section{Modified $f(T)$ cosmology}
\label{sec3}
Modified teleparallel models are obtained by replacing the torsion scalar $T$ in the action \eqref{T_Einstein} with a general function $f(T)$, and neglecting an explicit cosmological constant $\Lambda$. The aim is to find forms of $f(T)$ that drive late-time cosmic acceleration without $\Lambda$. Such extensions typically introduce extra degrees of freedom and lead to modified Friedmann equations. The choice of working with torsion instead of curvature is mainly due to its mathematical simplicity in modified models, since the field equations in $f(T)$ gravity remain second order, while in $f(R)$ gravity they become fourth order, making the teleparallel framework more tractable for analytical and numerical analysis \cite{Cai_2016}.\\

In general example we consider our modified action in Eq.(\ref{action_mod}) \cite{Sandoval_2024}
\begin{equation}
\label{action_mod}
    S = \frac{1}{2 \kappa^2}\int ed^{4}x [-T + f(T)] + \int ed^{4}x (\mathcal{L}_{m}+\mathcal{L}_{r}),
\end{equation}
the variation with respect to tetrad vector field, the field equation becomes
\begin{equation}
    \label{field_mod}
\begin{aligned}
e^{-1} D_\mu \Big( e S_a{}^{\mu\nu} \Big) f_T + S_a{}^{\mu\nu} (\partial_\mu T) f_{TT} + e^b{}_{\rho} T^{\rho}{}_{\mu a} S_b{}^{\nu\mu} f_T + \frac{1}{4} e_a{}^\nu f(T) = 4\pi G \, e_a{}^\rho \mathcal{T}^{\nu}_{\mu},
\end{aligned}
\end{equation}
where $f_{T}(T) = \partial f(T)/\partial T$, $f_{TT}(T) = \partial^{2}f(T)/\partial T^{2}$. It is important to emphasize that, when considering $f(T) = constant$, Eq.(\ref{action_mod}) will be reduced to the usual field Eq.(\ref{T_Einstein}) of teleparallelism equivalent to General Relativity. It is also relevant to remember that the solution of Eq.(\ref{field_mod}) is directly linked to the choice of tetrads for the model, thus, an appropriate set of tetrads can provide an adequate solution.\\

Solving Eq.~\eqref{field_mod} for the tetrad in Eq.~\eqref{tetrad} yields the modified Friedmann equations with perfect-fluid components \cite{Sandoval_2024}.
\begin{equation}
\label{friedmann_mod1}
    3H^{2} + Tf_{T}(T) - \frac{f(T)}{2} = \kappa^{2} \rho_i,
\end{equation}
\begin{equation}
\label{friedmann_mod2}
     \dot{H}\left(1-f_{T}(T) -2Tf_{TT}(T)\right) = -\frac{\kappa^2}{2}(\rho_i + p_i).
\end{equation}
If the components of universe are approximately perfect fluids, we can take the extra terms and give the same treatment, however, with geometrical origin. The effective energy density and pressure of extra terms are given by equations (\ref{energyeff}) and (\ref{presseff}),
\begin{equation}
\label{energyeff}
    \rho_{geo} = -\frac{1}{2\kappa^{2}}(2Tf_{T}(T)- f(T)),
\end{equation}
\begin{equation}
\label{presseff}
    p_{geo}= -\frac{1}{2 \kappa^2}(-8T\dot{H}f_{TT}(T) - f_{T}(T)(-2T + 4\dot{H})-f(T)).
\end{equation}
Using the relation between $\rho$ and $p$ of the perfect fluid discussed earlier $p = \omega \rho$, we have that the effective state parameter for this perfect fluid will be \cite{Bamba_2012,Capozziello_2024}\\
\begin{equation}
    \omega_{geo} = \frac{-8T\dot{H}f_{TT}(T) - f_{T}(T)(-2T + 4\dot{H})-f(T)}{2Tf_{T}(T)- f(T)}.
\end{equation}

\subsection{Expansion of modified gravity}
\label{sec31}

The dimensionless Hubble's parameter $E(a)$ for flat $\Lambda CDM$ model is given by Eq.(\ref{dimensionless_lcdm})
\begin{equation}
\label{dimensionless_lcdm}
    E(a) = \sqrt{\Omega_{r0} a^{-4} + \Omega_{m0} a^{-3} + \Omega_\Lambda},
\end{equation}
however, for a modified model with dynamical dark energy we have the Eq.(\ref{dimensionless_mod}) \cite{Sabiee_2022}
\begin{equation}
    \label{dimensionless_mod}
    E(a) = \sqrt{\Omega_{r0} a^{-4} + \Omega_{m0} a^{-3} + \Omega_{f0} X(a)},
\end{equation}
where by the principle of completeness
\begin{equation}
    \Omega_{f0} = 1 - \Omega_{m0} - \Omega_{r0},
\end{equation}
and $X(a)$ is a dynamical term given by \cite{Sabiee_2022}:
\begin{equation}
\label{xa}
    X(a) = \frac{1}{T_0 \Omega_{f0}}(-f(T) + 2Tf_T(T)),
\end{equation}
where $T_0 = -6 H_0^2$ is the torsion scalar present in section \ref{sec2} when $z =0$. For a simple example, we can consider a power-law model that has already been used in several works~\cite{Sandoval_2024,Sabiee_2022,Briffa:2021nxg,Sandoval-Orozco:2023pit}, $f(T) = \alpha (-T)^{\gamma}$, where the condition for return to $\Lambda CDM$ is consider $\gamma = 0$.
Following the same procedure used in the cited works, we can write the parameter $\alpha$ as a function of the other parameters, where $\alpha = (6 H_{0}^2)^{1-2\gamma} \frac{\Omega_{f0}}{1-2\gamma}$~\cite{Sabiee_2022}. So we can rewrite Eq.(\ref{dimensionless_mod}) as
\begin{equation}
\label{dimensionless_mod_mod}
E^2(a,\gamma) = \Omega_{r0}a^{-4} + \Omega_{0m}a^{-3} + \Omega_{f0}E^{2\gamma}(a,\gamma).
\end{equation}

If we make the expansion in a power series around $\gamma = 0$, we can rewrite Eq.(\ref{dimensionless_mod_mod}) as
\begin{equation}
E^2(a,\gamma) \approx E^2(a,0) + \Omega_{f0}\ln(E^2(a,0))\,\gamma + ...,
\end{equation}
where $E(a,0)$ is the dimensionless Hubble parameter for the $\Lambda$CDM model. In this way, we can write the parameter $E(a,\gamma)$ for a modified model in terms of the well-known $E(a,0)$.

\subsection{New $f(T)$ model}
\label{sec32}

For this work, we propose a new function of the torsion scalar motivated by some hypotheses. Returning to Eq.(\ref{energyeff}), we know that if we want to obtain the energy density of the cosmological constant, we must consider $\rho = \text{constant}$, such as $\Lambda/3$ in Eq.(\ref{friedmann}). Following the idea of a sum of multiple power laws presented in \cite{Aquino:2025sdb}, we can formulate a new functional form for Eq.~(\ref{energyeff}) as
\begin{equation}
    \rho_{geo} = -\alpha \left( \frac{\ln(2)}{\ln(2-\gamma)} - \frac{\gamma~T}{2~T_0~\ln(2 - \gamma)} - \frac{(\gamma~T)^2 }{8~T_0^2~\ln(2 - \gamma)}  - ...\right),
    \label{series}
\end{equation}
and the series of Eq.(\ref{series}) can be condensed in
\begin{equation}
    \label{rho_geo}
    \rho_{geo} = -\alpha~log_{2+\gamma}(2+\gamma~T/T_0),
\end{equation}
where $\alpha$ and $\gamma$ are constant parameters of model. If we assume $\gamma = 0$, the Eq.(\ref{rho_geo}) reduces to a constant value $\rho_{geo} = \alpha$ and we return to a cosmological constant idea.

Solving the diferential Eq.(\ref{energyeff}) using the relation in Eq.(\ref{rho_geo}) and the initial condition $f(T_0) = \alpha$ to recover the $\Lambda CDM$ model when $z = 0$, we have the function of torsion scalar $f(T)$ like
\begin{equation}
\label{ft}
    f(T) = \frac{\alpha \left(6 \sqrt{2 T \gamma} \left(\arctan\left(\sqrt{\frac{\gamma}{2}}\right) - \arctan\left(\sqrt{\frac{T \gamma}{2 T_0}}\right)\right) - 5 \sqrt{T} \ln(2 + \gamma) + 6 \sqrt{T_0} \ln\left(2 + \frac{T \gamma}{T_0}\right)\right)}{\sqrt{T_0} \ln(2 + \gamma)},
\end{equation}
Solving Eq.(\ref{friedmann_mod1}) for $T = T_0$ and $z =0$ we can rewrite $\alpha$ as
\begin{equation}
        \alpha = 6 H_0^{2}(1 -\Omega_{0r}-\Omega_{0m}).
\end{equation}
Using this new $\alpha$ and the function in Eq.(\ref{ft}) in Eq.(\ref{xa}) we have
\begin{equation}
\label{xa1}
    X(a) =-\frac{\ln\left(2 - \frac{H^2 \gamma}{H_0^2}\right) (1 - \Omega_{0m} - \Omega_{0r})}{\Omega_t \ln(2 - \gamma)},
\end{equation}
and finally using the equation Eq.(\ref{dimensionless_mod}) for result for our $X(a)$ in Eq.(\ref{xa1}) and making the expansion around condition to return to $\Lambda CDM$ model ($\gamma = 0$) the result for dimensionless modified Hubble parameter is
\begin{equation}
\label{mod_ez}
    E^{2}(a,\gamma) = E^{2}(a,0) + \frac{\gamma (E^{2}(a,0) -1)(1- \Omega_{0m} - \Omega_{0r}) }{2~\ln{(2)}}+ O[\gamma]^2.
\end{equation}

We restrict the analysis to $\gamma\ge0$. For $\gamma>0$ the argument of the logarithm in Eq.~(\ref{rho_geo}), $2+\gamma T/T_0$, is positive for all $T/T_0\ge1$, so the model is well defined from $z=0$ up to the CMB epoch. For $\gamma<0$ the argument vanishes at $|\gamma|H^2/H_0^2=2$, which limits the model to a finite redshift range. As a consequence, our constraints on $\gamma$ are reported as upper limits when the posterior peaks at the boundary $\gamma=0$.

Mathematically, the logarithm in Eq.~(\ref{rho_geo}) is well defined over the whole range $T/T_0\ge1$ only for $\gamma\ge0$: for $\gamma<0$ its argument, $2+\gamma T/T_0$, vanishes at $|\gamma|H^2/H_0^2=2$, which limits the exact model to a finite redshift range. Nevertheless, in the MCMC analysis we do not impose $\gamma\ge0$ as a hard prior. Since the model is used through its expansion around $\gamma=0$, Eq.~(\ref{mod_ez}), small negative values of $\gamma$ remain well defined over the entire redshift range probed by the data, and we chose to let the statistics decide the preferred sign instead of truncating the posterior at the boundary $\gamma=0$. We therefore regard negative values as acceptable as long as $|\gamma|\ll1$, so that the first-order expansion remains valid and the argument of the logarithm stays positive. Consequently, constraints on $\gamma$ are quoted as two-sided intervals.

\section{Observational data and treatment}
\label{sec4}
In this section, we evaluate the theoretical results obtained in the previous section (\ref{sec3}), specifically in subsection \ref{sec32}, for the modified $f(T)$ function given in Eq.~(\ref{ft}). We perform a Markov Chain Monte Carlo (MCMC) analysis using several datasets in order to estimate the cosmological parameters.Our analysis will contemplate the $H(z)$ measurements from Cosmic Chronometers \cite{Sharov}, three Type Ia Supernova datasets, DES-SN5YR \cite{DES_2024}, Pantheon+ \cite{Scolnic_2022}, and Union~3.0 \cite{Rubin_2025}, BAO measurements from DESI~2024 \cite{DES_2024}, and CMB measurements from Planck~2018 \cite{Planck_2018_params}. The numerical analysis was performed in Python through the maximization of the likelihood function $\mathcal{L}$, the MCMC sampling was carried out using the emcee library \cite{ForemanMackey2013}, and Getdist library for plots \cite{Lewis2019GetDist}.

\subsection{Local observations (low z)}
\label{sec41}
In this subsection, we summarize the local measurements employed in our analysis. These datasets provide direct, low-redshift information about the cosmic expansion history and serve as independent constraints on the modified $f(T)$ model.

\subsubsection{Cosmic chronometers}
These measurements of the $H(z)$ parameter come from systems of passively evolving galaxies in galaxy clusters, where the absence of significant interactions allows small differences in redshift to be associated with age estimates. The spectra of these galaxies are used to obtain the quantity $dt/dz$, which relates the differential age to the redshift variation. The used dataset has 58 data points with redshift ranging from $0.07 > z > 2.40$ \cite{Sharov}. The likelihood is given by
\begin{equation}
\label{likelihoodhz}
    \mathcal{L}_{H(z)} \propto e^{- \frac{\chi^2_{H(z)}}{2} },
\end{equation}
where
\begin{equation}
    \chi^2_{H(z)} = \sum_{i=1}^{58} \frac{\left[ H_{\mathrm{theo}}(z_i, \Theta) - H_{\mathrm{obs}}(z_i) \right]^2}{\sigma_i^2}.
\end{equation}

\begin{figure}
\centering
\includegraphics[width=0.99\linewidth]{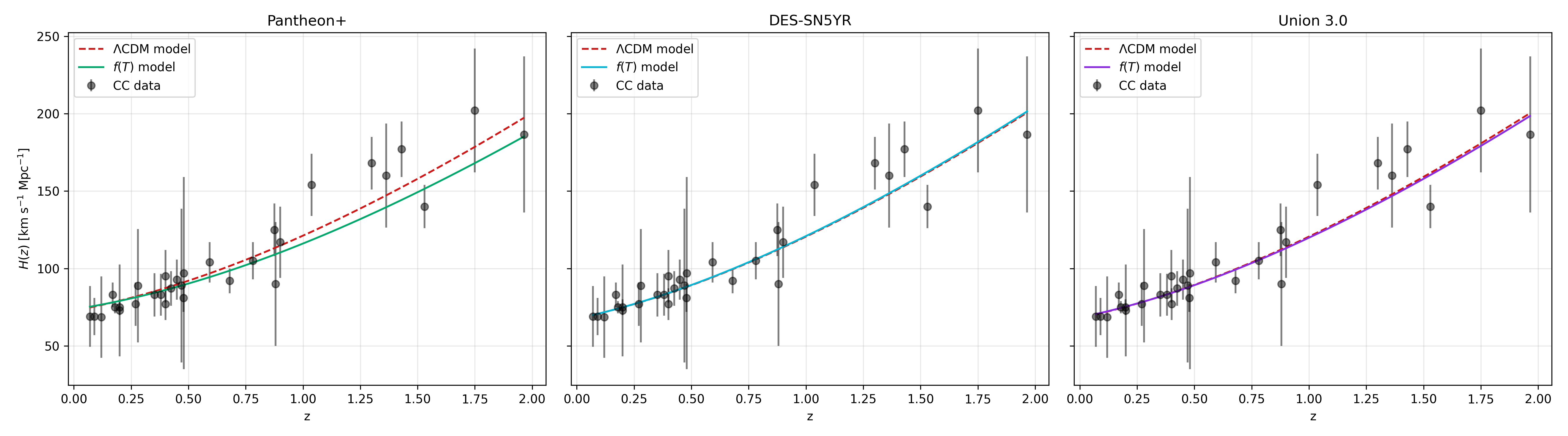}
\caption{Theoretical predictions of the $f(T)$ gravity model (solid lines) and standard $\Lambda\text{CDM}$ (dashed lines) contrasted against Cosmic Chronometers (CC) observational data points with 1-sigma error bars, broken down across Pantheon+, DES-SN5YR, and Union 3.0 dataset calibrations.}
\label{fig:hz}
\end{figure}

\subsubsection{Type Ia supernova dataset}

As mentioned earlier, we worked with three datasets: the DES-SN5YR \cite{DES_2024} sample, comprising 1827 light curves ($0.025< z < 1.13$); the Pantheon+ \cite{Scolnic_2022} compilation, containing 1701 light curves ($0.01 < z < 2.3$); and the Union~3.0 \cite{Rubin_2025} dataset, which includes 2087 light curves summarized into 22 redshift bins ($0.05 < z < 2.26$). In these datasets we used the distance modulus $\mu$ to measure the cosmological parameters. The distance modulus is given by \cite{}
\begin{equation}
    \mu(z,\Theta) = 5~log_{10}(d_{L}(z,\Theta)/ 1 Mpc) + 25,
    \label{distance_modulus}
\end{equation}
where $d_{L}$ is the luminosity distance given by
\begin{equation}
    d_{L}(z,\Theta) = c~\frac{(1+z)}{H_0}~\int_{0}^z \frac{dz'}{E(z',\Theta)}.
\end{equation}
The likelihood for Ia supernova data is
\begin{equation}
\label{likelihoodsne}
    \mathcal{L}_{H(z)} \propto e^{- \frac{\chi^2_{SN}}{2} },
\end{equation}
where
\begin{equation}
    \chi_{SN}^2 = \Delta\mu_i~C^{-1}_{ij}~\Delta\mu_j,
\end{equation}
and $\Delta\mu$ is the difference between the theoretical distance modulus and the measured one,
$\Delta\mu = \mu(z,\Theta) - \mu_{\mathrm{obs}}$, and $C^{-1}_{ij}$ is the inverse covariance matrix that contains the measurement errors of $\mu$.

\begin{center}
\includegraphics[width=0.99\linewidth]{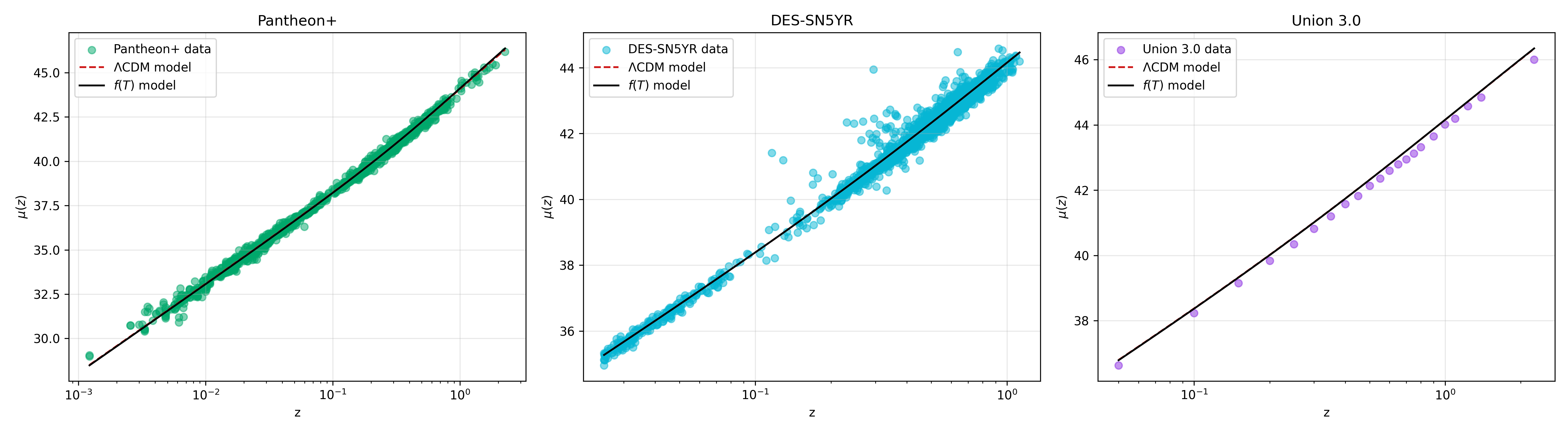}
\captionof{figure}{Comparison between theoretical distance modulus curves and observed supernova binned data ($\mu(z)$) across Pantheon+, DES-SN5YR, and Union 3.0 samples plotted against redshift in logarithmic scale.}
\label{fig:mu}
\end{center}

\subsubsection{Baryon Acoustic Oscillation (BAO) data}

The DESI DR2 BAO measurements \cite{Adame2025DESI} provide some of the most precise constraints on the late-time expansion history of the Universe. The dataset covers the redshift range $0.3 \lesssim z_{\mathrm{eff}} \lesssim 2.33$, with measurements obtained from different tracers. Table~\ref{tb1} summarizes the observables employed in this work: the transverse comoving distance $D_{M}/r_{d}$, the radial BAO scale $D_{H}/r_{d}$, and, depending on the tracer, either the distortion parameter $r$ or the volume-averaged distance $D_{V}/r_{d}$. All values are normalized by the sound horizon at the drag epoch, $r_{d}$, computed consistently for each cosmological model.\\

The theoretical predictions for the BAO quantities are derived from the cosmological expansion history. For a homogeneous and isotropic FLRW universe, we use
\begin{equation}
    D_{M}(z) = (1+z)\,D_{A}(z),
\end{equation}
\begin{equation}
    D_{H}(z) = \frac{c}{H(z)},
\end{equation}
\begin{equation}
    D_{V}(z) =
    \left[
        z\, D_M^2(z)\,\frac{c}{H(z)}
    \right]^{1/3},
\end{equation}
where $D_A = \frac{c}{(1+z)} \int_0^z \frac{dz'}{H(z')}$. The comparison with DESI is performed through the normalized ratios
\begin{equation}
\frac{D_M(z)}{r_d}, \qquad
\frac{D_H(z)}{r_d}, \qquad
\frac{D_V(z)}{r_d},
\end{equation}
evaluated at the effective redshift of each tracer. For the sound horizon at the drag epoch we adopt the DESI DR2 fitting formula \cite{Adame2025DESI}
\begin{equation}
    r_d = 147.05
    \left( \frac{\omega_m}{0.1432} \right)^{-0.23}
    \left( \frac{N_{\mathrm{eff}}}{3.04} \right)^{-0.10}
    \left( \frac{\omega_b}{0.02236} \right)^{-0.13}
    \ \mathrm{Mpc},
\end{equation}
where we used $N_{eff} = 3.046$ in accordance with the standard particle model \cite{PDG2024}, $\omega_b = \Omega_bh^2$ and $\omega_m = \Omega_mh^2$.\\

Since we do not use covariance matrices in this work, each BAO measurement contributes independently to the likelihood. For each observable $\mathcal{D}$ in Table~\ref{tb1}, we define the residual
\begin{equation}
\Delta\mathcal{D}_i =
\mathcal{D}_{\mathrm{th}}(z_i,\Theta)
-
\mathcal{D}_{\mathrm{obs}}(z_i),
\end{equation}
and compute the chi-square and likelihood assuming uncorrelated Gaussian uncertainties:
\begin{equation}
\chi^2_{\mathrm{BAO}}(\Theta)
=
\sum_{i}
\frac{\big[\Delta\mathcal{D}_i\big]^2}{\sigma_i^2},
\end{equation}
\begin{equation}
\label{likelihoodbao}
    \mathcal{L}_{BAO} \propto e^{-\frac{\chi^2_{BAO}}{2}}
\end{equation}

Although this simplified treatment does not incorporate possible correlations between transverse and radial BAO modes, it nonetheless captures the main constraining power of the DESI DR2 sample and yields reliable estimates of the expansion history encoded in the BAO observables.

\begin{table}[t]
\centering
\small
\caption{DESI BAO measurements DR2 [https://arxiv.org/pdf/2503.14738], including 1D isotropic ($D_V/r_{\mathrm{d}}$) and 2D anisotropic ($D_M/r_{\mathrm{d}}, D_H/r_{\mathrm{d}}$) observables with their correlation coefficient $r_{\mathrm{M,H}}$.}
\label{tb1}
\begin{tabular}{lccccc}
\toprule
Tracer & $z_{\mathrm{eff}}$ & $D_V/r_{\mathrm{d}}$ & $D_M/r_{\mathrm{d}}$ & $D_H/r_{\mathrm{d}}$ & $r_{\mathrm{M,H}}$ \\
\midrule
BGS         & 0.295 & $7.942 \pm 0.075$  & ---                & ---                & ---      \\
LRG1        & 0.510 & ---                & $13.588 \pm 0.167$ & $21.863 \pm 0.425$ & $-0.459$ \\
LRG2        & 0.706 & ---                & $17.351 \pm 0.177$ & $19.455 \pm 0.330$ & $-0.404$ \\
LRG3 + ELG1 & 0.934 & ---                & $21.576 \pm 0.152$ & $17.641 \pm 0.193$ & $-0.416$ \\
ELG2        & 1.321 & ---                & $27.601 \pm 0.318$ & $14.176 \pm 0.221$ & $-0.434$ \\
QSO         & 1.484 & ---                & $30.512 \pm 0.760$ & $12.817 \pm 0.516$ & $-0.500$ \\
Lya         & 2.330 & ---                & $38.988 \pm 0.531$ & $8.632 \pm 0.101$  & $-0.431$ \\
\midrule
LRG3        & 0.922 & ---                & $21.648 \pm 0.178$ & $17.577 \pm 0.213$ & $-0.406$ \\
ELG1        & 0.955 & ---                & $21.707 \pm 0.335$ & $17.803 \pm 0.297$ & $-0.462$ \\
\bottomrule
\end{tabular}
\end{table}

\subsection{Global observations (high z)}
\label{sec42}
In this subsection, we summarize the high-redshift measurements employed in our analysis. These datasets probe the cosmic expansion history at intermediate and distant epochs, providing information that complements the low-redshift observations discussed previously. High-redshift data are particularly valuable because they capture the behavior of the Universe when matter dominated the expansion, allowing us to test the consistency of the modified $f(T)$ model over a wide dynamical range.\\

By extending the observational window to redshifts well above unity, these measurements help constrain the model in regimes where deviations from standard cosmology may become more apparent. They also ensure that the background evolution remains physically consistent from early times to the present day, strengthening the overall robustness of our parameter inferences.\\

\subsubsection{Cosmic Background Microwave (CMB)}
For high redshift we choose the CMB data from Planck 18 \cite{Planck_2018_params} distance priors. This dataset can provide information on the CMB power spectrum without the calculation of the complete set of perturbations \cite{Chen2019DistancePriors}. In this dataset we have the following quantities: The shift parameter $R$ that measures the peak spacing of the temperature in the
power spectrum, the combination $\omega_{b}$ and the acoustic scale $l_A$ measures the temperature in the transverse direction.\\

Those quantities are represented in a vector $V = \left\{ R, l_A, \omega_b \right\}$ and respectively given by Eqs.(\ref{shift}) and (\ref{acoustic})\\
\begin{equation}
\label{shift}
    R(z) = \frac{\sqrt{\Omega_{m}H_0}(1+z)D_A(z)}{c},
\end{equation}
\begin{equation}
\label{acoustic}
    l_A (z) = \frac{\pi (1+z)D_A(z)}{r_s(z)},
\end{equation}
and $\omega_b$ is a value given by Planck 18. The quantities of angular diameter distance $D_A$ and comoving sound horizon $r_s$ are defined by\\
\begin{equation}
    D_A (z) = \frac{c}{(1+z)}\int^z_0\frac{dz'}{H(z')}
\end{equation}
and
\begin{equation}
    r_s(z) = c \int^{1/(1+z)}_0 \frac{da}{a^2H(a)\sqrt{1+a\frac{3\omega_b}{4\omega_\gamma}}},
\end{equation}
where $\omega_\gamma$ it is related to the fractional density of photons, and  the term $\frac{3}{4\omega_\gamma} = 31500(T_{CMB}/2.7K)^{-4}$ \cite{Chen2019DistancePriors}. For our analysis we use $\Omega_r = \frac{\Omega_m}{1+z_{eq}}$, where $z_{eq} = 2.5 \times 10^4 \omega_m (T_{CMB}/2.7K)^{-4}$ \cite{Chen2019DistancePriors}.\\

The chi-square and likelihhod function of Planck 18 is given by\\
\begin{equation}
    \chi^2_{CMB} = \Delta V_i C^{-1}_{ij~CMB}\Delta V_j,
\end{equation}
\begin{equation}
\label{likelihoodcmb}
    \mathcal{L}_{CMB} \propto e^{-\frac{\chi^2_{CMB}}{2}}
\end{equation}
where $\Delta V(\Theta)$ is the difference between the observed values of the quantities and their theoretical predictions $\Delta V = V(\Theta) - V $, where the vector $V(\Theta) = [R, l_A , \omega_b]$ and $C_{\mathrm{CMB}}$ is the covariance matrix of the corresponding uncertainties. The matrix $C_{\mathrm{CMB}}$ is constructed from the correlation matrix $c_{ij}$, given in Eq.~(\ref{cij}), together with the standard deviations $\sigma_i$ and $\sigma_j$ associated with the quantities $R$, $l_A$, and $\Omega_b$
\cite{Sandoval_2024}.\\
\begin{equation}
\label{cij}
    c_{ij} =\begin{pmatrix}
1 &0.46  &-0.66  \\
0.46 & 1 & -0.33 \\
 -0.66& -0.33 & 1 \\
\end{pmatrix}.
\end{equation}\\
The values and standard deviations from Planck 18 are $R = 1.7502 \pm 0.0046$, $l_A = 301.471 \pm 0.090$, and $\omega_b = 0.02236 \pm 0.00015$~\cite{Sandoval_2024,Planck_2018_params}. In this way, $\boldsymbol{\sigma}_i$ and $\boldsymbol{\sigma}_j$ denote, respectively, the column and row vectors of the $1\sigma$ uncertainties associated with the quantities $R$, $l_A$, and $\Omega_b$. finaly, the covariance matrix is defined as\\
\begin{equation}
\label{cijcmb}
    C_{ij~CMB} = c_{ij}\sigma_i\sigma_j.
\end{equation}\\
Making the inverse matrix of $C_{ij~CMB}$ we have the Table \ref{cmbinvcov}.\\

\begin{table}[t]
\centering
\begin{tabular}{r r r}
  $9.47889969 \times 10^{4}$  & $-1.31680342 \times 10^{3}$ & $1.65780222 \times 10^{6}$ \\
  $-1.31680342 \times 10^{3}$ & $1.56837219 \times 10^{2}$  & $4.40166823 \times 10^{3}$ \\
  $1.65780222 \times 10^{6}$  & $4.40166823 \times 10^{3}$  & $7.88698917 \times 10^{7}$ \\
\end{tabular}
\caption{Covariance matrix $C_{ij ~ CMB}$ built based on Eq.~(\ref{cijcmb}).}
\label{cmbinvcov}
\end{table}

\section{Results}
\label{sec5}
This section presents the results for the modification introduced in Eq.~(\ref{ft}), using the data presented in Section~\ref{sec4}. The results are obtained by maximizing the likelihood functions given in Eqs.~(\ref{likelihoodhz}), (\ref{likelihoodsne}), (\ref{likelihoodbao}), and (\ref{likelihoodcmb}) through an MCMC procedure implemented with the emcee library \cite{ForemanMackey2013}.

\begin{table}[H]
\centering
\renewcommand{\arraystretch}{1.5}
\setlength{\tabcolsep}{4pt}
\small
\caption{Best-fit parameters for the $\Lambda$CDM and $f(T)$ models using Cosmic Chronometers (CC), DESI DR 2, and Planck 2018 CMB data combined with three different supernova samples: (I) DES-SN5YR, (II) Pantheon+SH0ES, and (III) Union 3.0. The reported uncertainties correspond to $1\sigma$ uncertainties.}
\label{tab:models_comparison}
\begin{tabular}{c c c c c c c}
\hline\hline
Attempt & Model & $\Omega_m$ & $H_0$ & $\Omega_b$ & $\gamma$ & $\chi^2_{\min}$ \\
& & & (km s$^{-1}$ Mpc$^{-1}$) & & & \\
\hline
I & $\Lambda$CDM & $0.3112^{+0.0041}_{-0.0042}$ & $67.6391^{+0.3186}_{-0.3223}$ & $0.0490^{+0.0004}_{-0.0004}$ & --- & 1680.155 \\
  & $f(T)$       & $0.3123^{+0.0046}_{-0.0045}$ & $67.6578^{+0.3261}_{-0.3210}$ & $0.0491^{+0.0004}_{-0.0004}$ & $0.0092^{+0.0127}_{-0.0129}$ & 1679.631\\
\hline
II & $\Lambda$CDM & $0.2533^{+0.0022}_{-0.0023}$ & $72.7756^{+0.2206}_{-0.2056}$ & $0.0442^{+0.0003}_{-0.0003}$ & --- & 683.551 \\
   & $f(T)$       & $0.2276^{+0.0041}_{-0.0036}$ & $73.5915^{+0.2263}_{-0.2413}$ & $0.0420^{+0.0004}_{-0.0004}$ & $-0.1190^{+0.0158}_{-0.0152}$ & 634.211 \\
\hline
III & $\Lambda$CDM & $0.3059^{+0.0056}_{-0.0052}$ & $68.0295^{+0.4173}_{-0.4247}$ & $0.0486^{+0.0005}_{-0.0005}$ & --- & 60.560 \\
    & $f(T)$      & $0.3005^{+0.0097}_{-0.0096}$ & $68.2777^{+0.5482}_{-0.5575}$ & $0.0481^{+0.0008}_{-0.0008}$ & $-0.0146^{+0.0215}_{-0.0210}$ & 59.944 \\
\hline\hline
\end{tabular}
\end{table}

\begin{figure}
    \centering
    \includegraphics[width=0.6\linewidth]{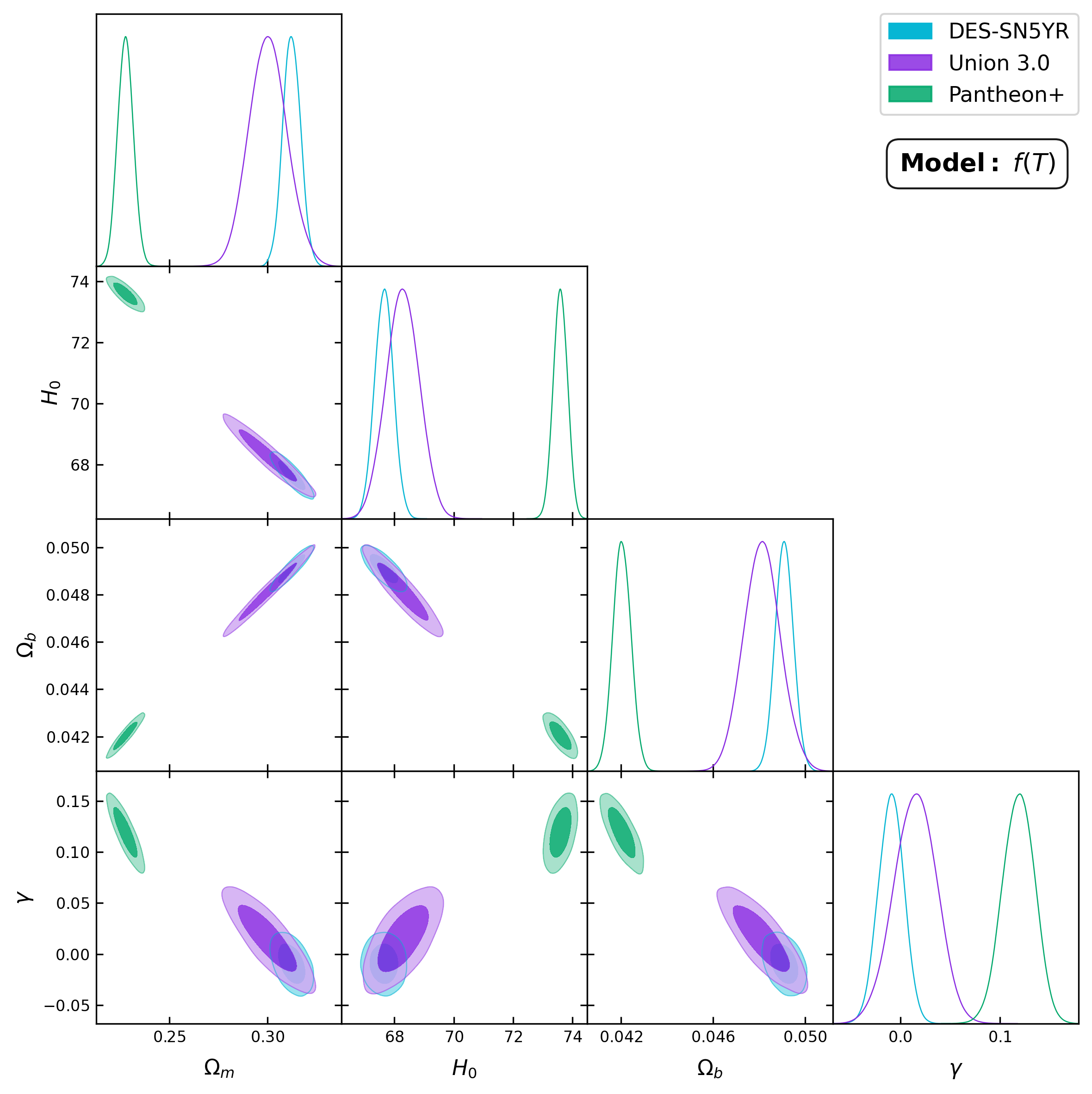}
    \caption{Caption}
    \label{fig:corner_mod}
\end{figure}

\begin{figure}
    \centering
    \includegraphics[width=0.6\linewidth]{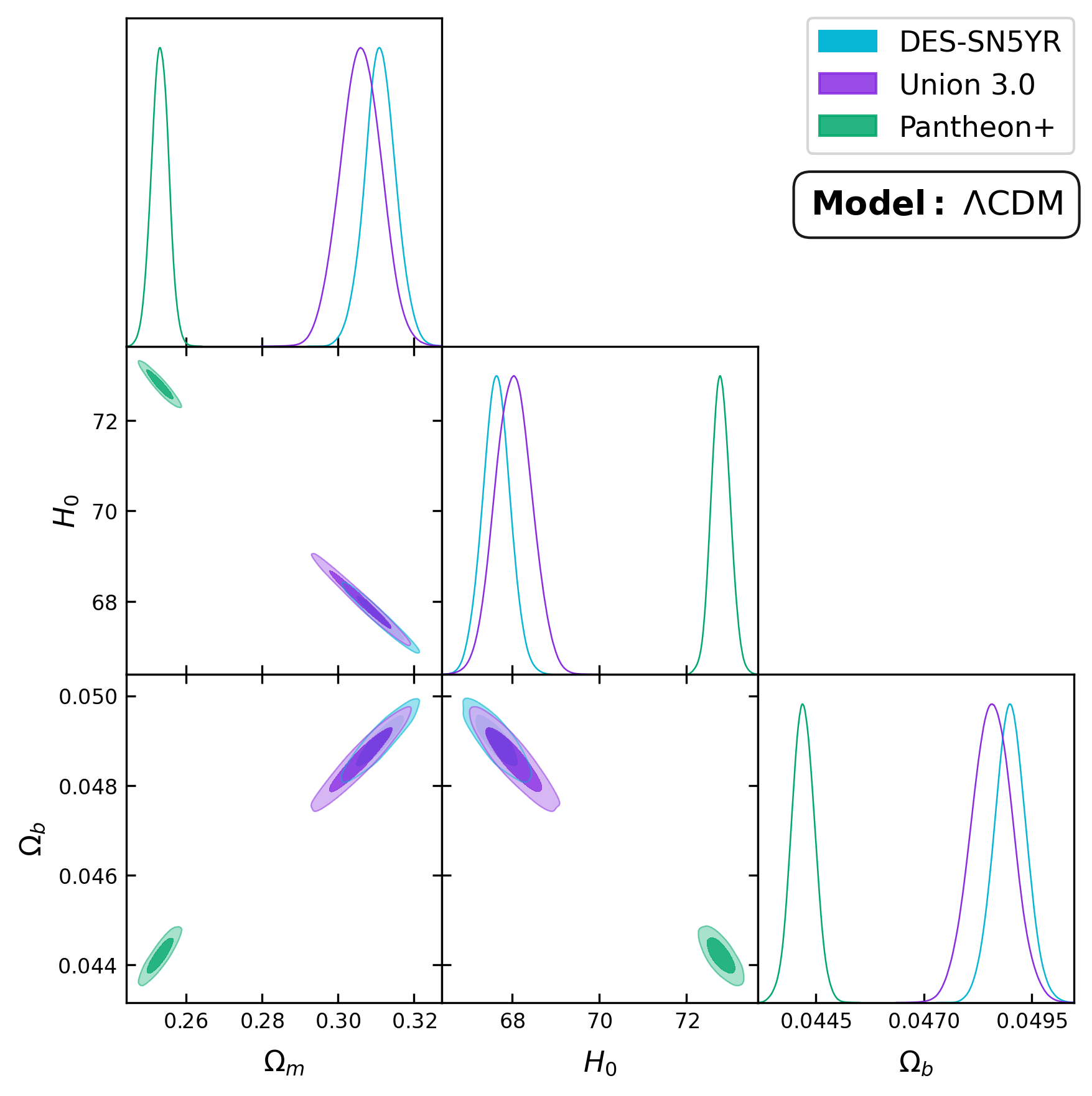}
    \caption{MCMC posterior distributions and 1$\sigma$--2$\sigma$ confidence contours for the standard $\Lambda\text{CDM}$ model parameters ($\Omega_m$, $H_0$, $\Omega_b$), contrasted across DES-SN5YR, Union 3.0, and Pantheon+ supernova samples.}
    \label{fig:corner_lcdm}
\end{figure}

The analysis was performed in three stages, one for each different supernova dataset, while keeping all other remaining data. For comparison, we also ran the modified gravity model $f(T)$ and $\Lambda CDM$ model using the same datasets and parameter sets. The results are shown in Table \ref{tab:models_comparison}. By comparing the corresponding parameters, we find that the values are very similar and agree within the $1\sigma$ uncertainties. These results are important for our hypothesis, as they show that we can reproduce cosmological-constant--like behavior without introducing an explicit cosmological constant, but instead through a dynamical function of the torsion scalar.\\

\begin{table}[H]
\centering
\renewcommand{\arraystretch}{2.0}
\setlength{\tabcolsep}{4pt}
\small
\caption{Best-fit parameters for the $\Lambda$CDM and $f(T)$ models using Planck 2018 CMB data alone and combined progressively with DESI DR2 BAO and Cosmic Chronometers (CC), without supernova data. The reported uncertainties correspond to $1\sigma$ uncertainties.}
\label{tab:models_probes}
\begin{tabular}{l l c c c c c}
\hline\hline
Dataset & Model & $\Omega_m$ & $H_0$ & $\Omega_b$ & $\gamma$ & $\chi^2_{\min}$ \\
& & & (km s$^{-1}$ Mpc$^{-1}$) & & & \\
\hline
CMB & $f(T)$ & $0.376^{+0.081}_{-0.107}$ & $64.311^{+5.770}_{-3.293}$ & $0.054^{+0.006}_{-0.008}$ & $0.094^{+0.131}_{-0.170}$ & $0.01$ \\
 & $\Lambda\text{CDM}$ & $0.3171_{-0.0083}^{+0.0086}$ & $67.1981_{-0.6305}^{+0.6194}$ & $0.0495_{-0.0007}^{+0.0007}$ & --- & $0.004$ \\
\hline
CMB + BAO & $f(T)$ & $0.2847^{+0.0107}_{-0.0100}$ & $69.0993^{+0.6162}_{-0.6127}$ & $0.0469^{+0.0009}_{-0.0008}$ & $-0.0479^{+0.0233}_{-0.0224}$ & $12.866 $ \\
 & $\Lambda\text{CDM}$ & $0.3036_{-0.0055}^{+0.0054}$ & $68.2124_{-0.4285}^{+0.4374}$ & $0.0484_{-0.0005}^{+0.0005}$ & --- & $17.193$ \\
\hline
CMB + BAO + CC & $f(T)$ & $0.2872^{+0.0104}_{-0.0098}$ & $68.9819^{+0.5952}_{-0.6017}$ & $0.0471^{+0.0009}_{-0.0008}$ & $-0.0414^{+0.0227}_{-0.0225}$ & $28.245$ \\
 & $\Lambda\text{CDM}$ & $0.3036_{-0.0057}^{+0.0055}$ & $68.2154_{-0.4354}^{+0.4443}$ & $0.0484_{-0.0005}^{+0.0005}$ & --- & $31.931$ \\
\hline\hline
\end{tabular}
\end{table}

To assess how each observational probe contributes to the constraints, we also performed the analysis without supernova data, adding the datasets progressively: Planck 2018 CMB alone, CMB combined with DESI DR2 BAO, and CMB + BAO + CC. The results are presented in Table~\ref{tab:models_probes}. Since the CMB information is compressed into only three data points ($R$, $\ell_A$, and $\omega_b$), the CMB-only fit has no statistical power to discriminate between the models: the $\Lambda$CDM model has three free parameters and zero degrees of freedom, while the $f(T)$ model, with four free parameters, is underdetermined. Consequently, $\chi^2_{\min}\approx0$ in both cases and cannot be used for model comparison, and the parameters of the $f(T)$ model are only weakly constrained, with strong degeneracies among $\Omega_m$, $H_0$, and $\gamma$ ($\Omega_m=0.376^{+0.081}_{-0.107}$, $H_0 = 64.311^{+5.770}_{-3.293}$ km s$^{-1}$ Mpc$^{-1}$ and $\gamma = 0.094^{+0.131}_{-0.170}$, consistent with $\gamma = 0$ within $1\sigma$). Including BAO substantially tightens the constraints and breaks the degeneracy, and $\gamma$ becomes $-0.0479^{+0.0233}_{-0.0224}$, which deviates from zero at about $2.1\sigma$, while the $f(T)$ model attains a lower $\chi^2_{\min}$ than $\Lambda$CDM ($12.866$ versus $17.193$). The change of sign of $\gamma$ with respect to the CMB-only case is not significant, since in the latter $\gamma$ is compatible with zero, and reflects the degeneracy among $\Omega_m$, $H_0$ and $\gamma$ being lifted by the BAO data. Adding CC data produces essentially the same behavior ($\gamma=-0.0414^{+0.0227}_{-0.0225}$, $\sim1.8\sigma$ from zero, $\chi^2_{\min} = 28.245$ versus $31.931$), indicating that CC measurements are consistent with the CMB + BAO constraints but do not significantly improve them. When BAO data are included, $f(T)$ prefers a slightly lower $\Omega_m$ and a slightly higher $H_0$ than $\Lambda$CDM, although the shifts in $H_0$ remain within about $1.5\sigma$ and therefore do not resolve the Hubble tension by themselves.

The Confidence contours were calculated for both models and for all cited datasets. They are represented in Fig.(\ref{fig:corner_mod}) for modified $f(T)$ model and in Fig. (\ref{fig:corner_lcdm}) for $\Lambda CDM$ model.\\

After obtaining the best-fit cosmological parameters, we can analyze how the $f(T)$ model differs from the $\Lambda$CDM model through well-known derived quantities such as the Hubble parameter $H(z)$ and the deceleration parameter $q(z)$, defined by Eqs.~(\ref{hubble_param}) and (\ref{deceleration_param}), respectively:
\begin{equation}
    \label{hubble_param}
    H(z, \Omega_m, H_0, \gamma) = H_0 \sqrt{E(z, \Omega_m, H_0, \gamma)},
\end{equation}
\begin{equation}
    \label{deceleration_param}
    q(z, \Omega_m, H_0, \gamma) = -1 + \frac{(1+z)}{H(z, \Omega_m, H_0, \gamma)} \frac{dH(z, \Omega_m, H_0, \gamma)}{dz}.
\end{equation}
The Hubble parameter describes how fast the universe is expanding at a given moment in time. It quantifies the rate at which distances between galaxies increase due to the expansion of space itself, rather than through their own motion. The deceleration parameter $q(z)$ quantifies the acceleration of the cosmic expansion: the more negative $q(z)$ is, the more accelerated the expansion becomes. For plot these parameters, we use the Eq.(\ref{mod_ez}) for modified $f(T)$ model and assumes $\gamma = 0$ in the same equation for $\Lambda CDM$, and they are represented in Figures (\ref{q_mod}, \ref{q_lcdm}).

In the following analysis, we plot the $H(z)$ measurements derived from cosmic chronometers data in Fig. (\ref{fig:hz}), together with the distance modulus $\mu$ obtained from the DES-SN5YR, Pantheon+SH0ES, and Union~3.0 supernova datasets in Fig. (\ref{fig:mu}). The purpose is to directly compare the predictions of the standard cosmological model and the $f(T)$ gravity model against the observational data points. By confronting both models with the expansion history and supernova distance measurements, we assess how accurately each framework reproduces the observed behavior of the universe and investigate possible deviations from the standard cosmological scenario.

In all cases, we observed good agreement between the two models and the data. This reinforces our hypothesis of a perturbative dynamical function for energy density (see again the Eq.(\ref{rho_geo})) that could simulate the behavior of a cosmological constant.It is worth noting that, despite the fluid-like treatment, we are not dealing with a real physical fluid, since its origin is geometric rather than arising from a genuine source described by the energy--momentum tensor.

\begin{center}
\includegraphics[width=0.75\linewidth]{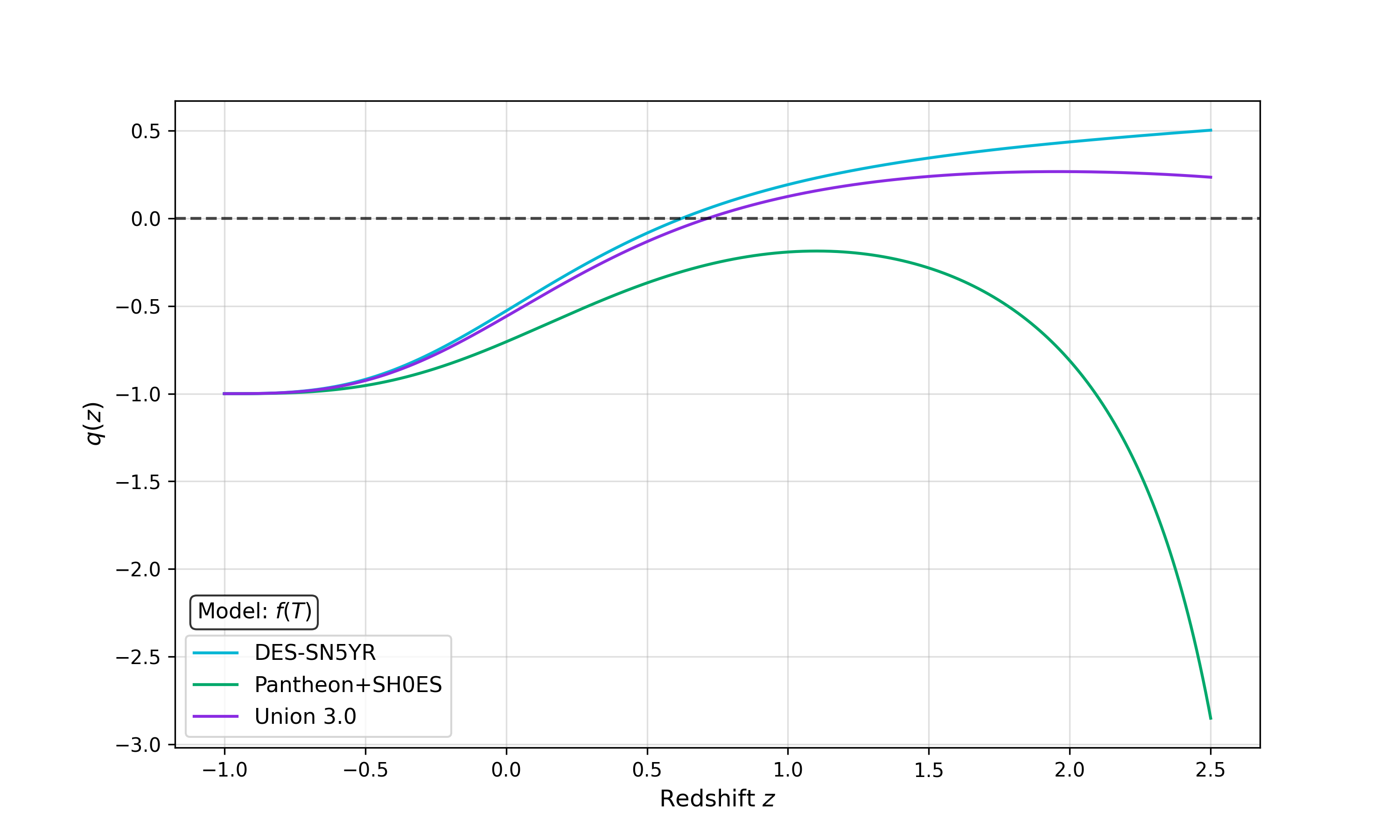}
\captionof{figure}{ Evolution of the deceleration parameter $q(z)$ for the $f(T)$ gravity model using diverse supernova datasets (DES-SN5YR, Pantheon+SH0ES, Union 3.0). Notice the dramatic divergence at high redshifts for one of the datasets, highlighting a potential tension.}
\label{q_mod}
\end{center}

\begin{center}
\includegraphics[width=0.75\linewidth]{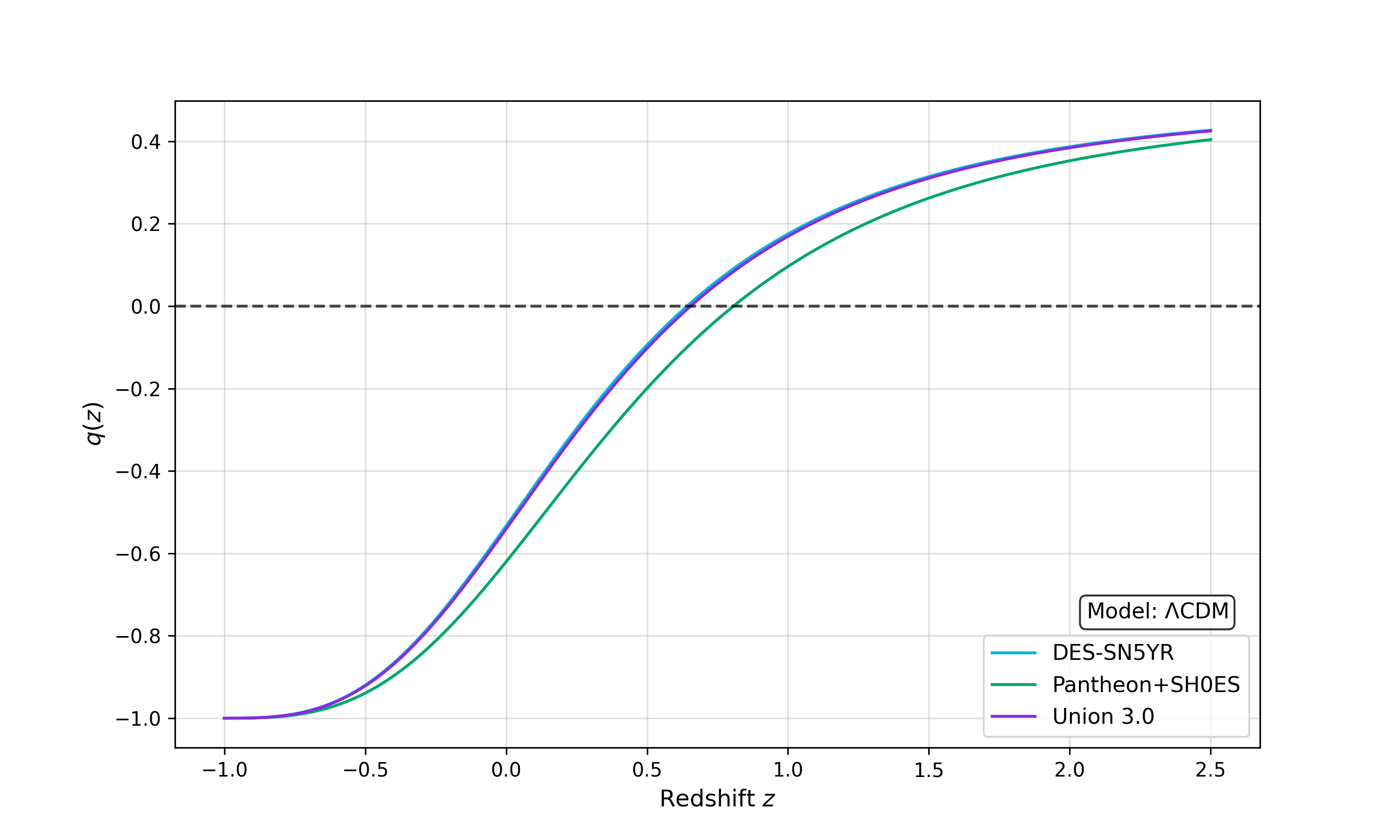}
\captionof{figure}{Evolution of $q(z)$ for the standard $\Lambda\text{CDM}$ model. Note the relative consistency and convergence of all supernova-based predictions, lacking the dramatic divergence seen in the $f(T)$ model.}
\label{q_lcdm}
\end{center}

Fig.(\ref{comparacao}) shows the marginalized constraints on the Hubble constant, $H_0$, and the present-day matter density parameter, $\Omega_m$, obtained for the $\Lambda$CDM and $f(T)$ gravity models using different combinations of observational datasets. In all cases, the analyses include DESI DR2, cosmic chronometers (CC), and Planck 2018 data, combined with distinct Type Ia supernova compilations, labeled as datasets [I], [II], and [III]. The left panel displays the inferred values of $H_0$, while the right panel shows the corresponding constraints on $\Omega_m$. The horizontal error bars represent the $1\sigma$ confidence intervals. The shaded vertical bands indicate reference regions for comparison: the red band corresponds to lower $H_0$ values typically associated with CMB-preferred estimates, whereas the blue band highlights higher $H_0$ values commonly linked to late-time measurements. This figure allows a direct comparison of how each cosmological model responds to changes in the supernova dataset, emphasizing shifts in the central values and uncertainties of $H_0$ and $\Omega_m$. In particular, it illustrates whether the $f(T)$ framework systematically alters the inferred parameter space relative to the standard $\Lambda$CDM scenario and whether such deviations may contribute to alleviating tensions in current cosmological constraints.

\begin{center}
  \includegraphics[width=0.8\linewidth]{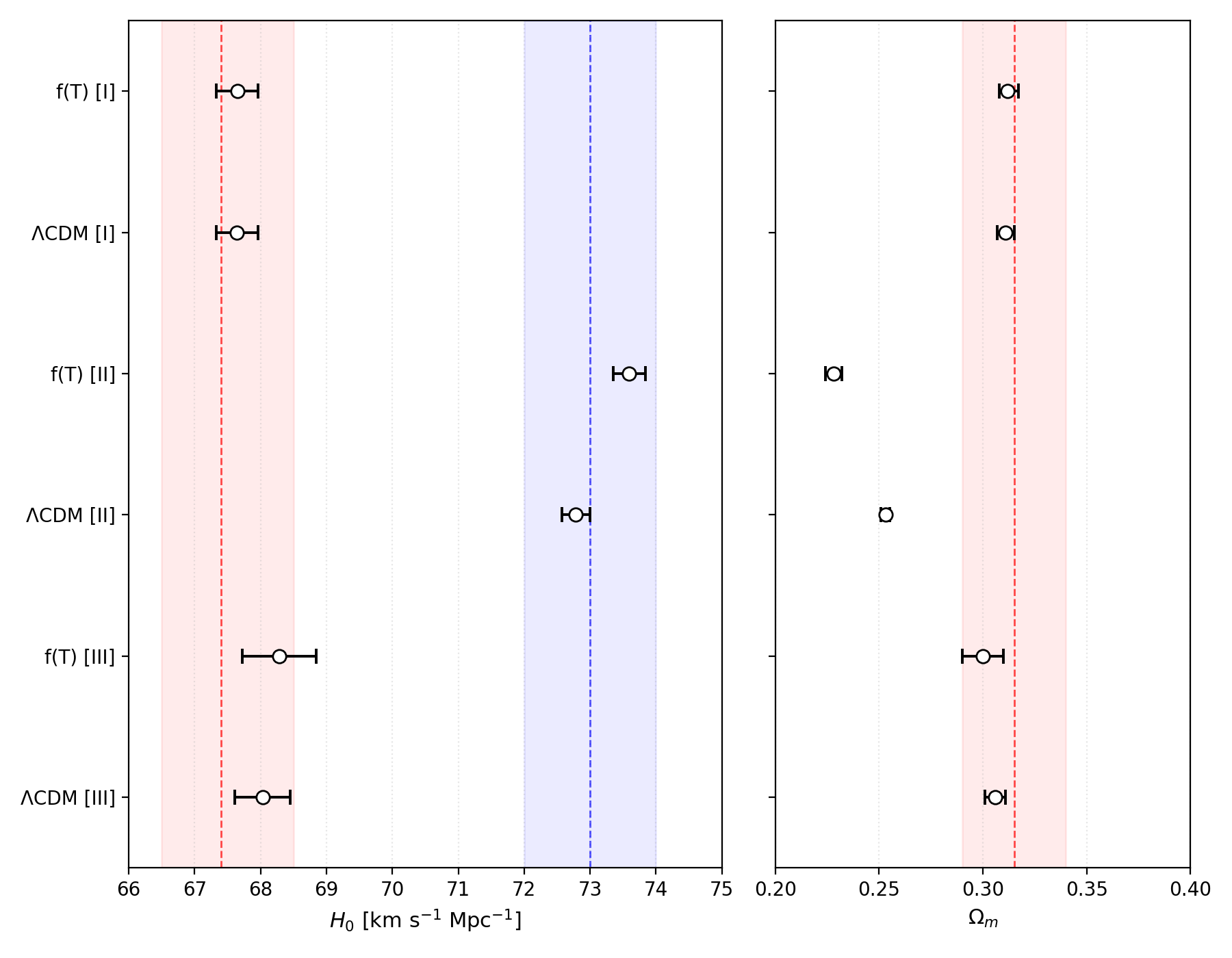}
  \captionof{figure}{Marginalized constraints on $H_0$ (left panel) and $\Omega_m$ (right panel) for the $\Lambda\text{CDM}$ and $f(T)$ models obtained from different combinations of datasets. All analyses include DESI DR2, cosmic chronometers (CC), and Planck 2018 data, combined with three distinct Type Ia supernova compilations, labeled as [I], [II], and [III]. The points represent the best-fit values, while the horizontal error bars correspond to the $1\sigma$ confidence intervals. The shaded vertical bands indicate reference regions for comparison, highlighting preferred ranges from early- and late-time measurements.}
  \label{comparacao}
\end{center}

\subsection{Implications for the Hubble tension}

To quantify whether the $f(T)$ model alleviates the Hubble tension, we compare the inferred $H_0$ with the SH0ES value $H_0 = 73.04\pm1.04$ km s$^{-1}$ Mpc$^{-1}$ \cite{Scolnic_2022}, adding the uncertainties in quadrature. For DES-SN5YR, the tension is $\sim 4.9\sigma$ for both models, $4.95\sigma$ for $f(T)$ and $4.96\sigma$ for $\Lambda$CDM). For Union~3.0, it is reduced only marginally, from $4.5\sigma$ ($\Lambda$CDM) to $4.0\sigma$ ($f(T)$. For CMB+BAO, the tension decreases from $4.3\sigma$ to $3.3\sigma$ with $f(T)$, and for CMB+BAO+CC from $4.3\sigma$ to $3.4\sigma$. These reductions arise mainly from the shift of $H_0$ by $\sim 1\sigma$ together with the larger uncertainties in the extra parameter, and none brings the tension below $3\sigma$. The only case in which $H_0$ is compatible with the local value is Pantheon+SH0ES, which already includes the SH0ES calibration; the agreement is therefore built into the data and is not an independent resolution of the tension. We conclude that the model, in its present form, mitigates but does not resolve the Hubble tension.

\begin{figure}
    \centering
    \includegraphics[width=0.7\linewidth]{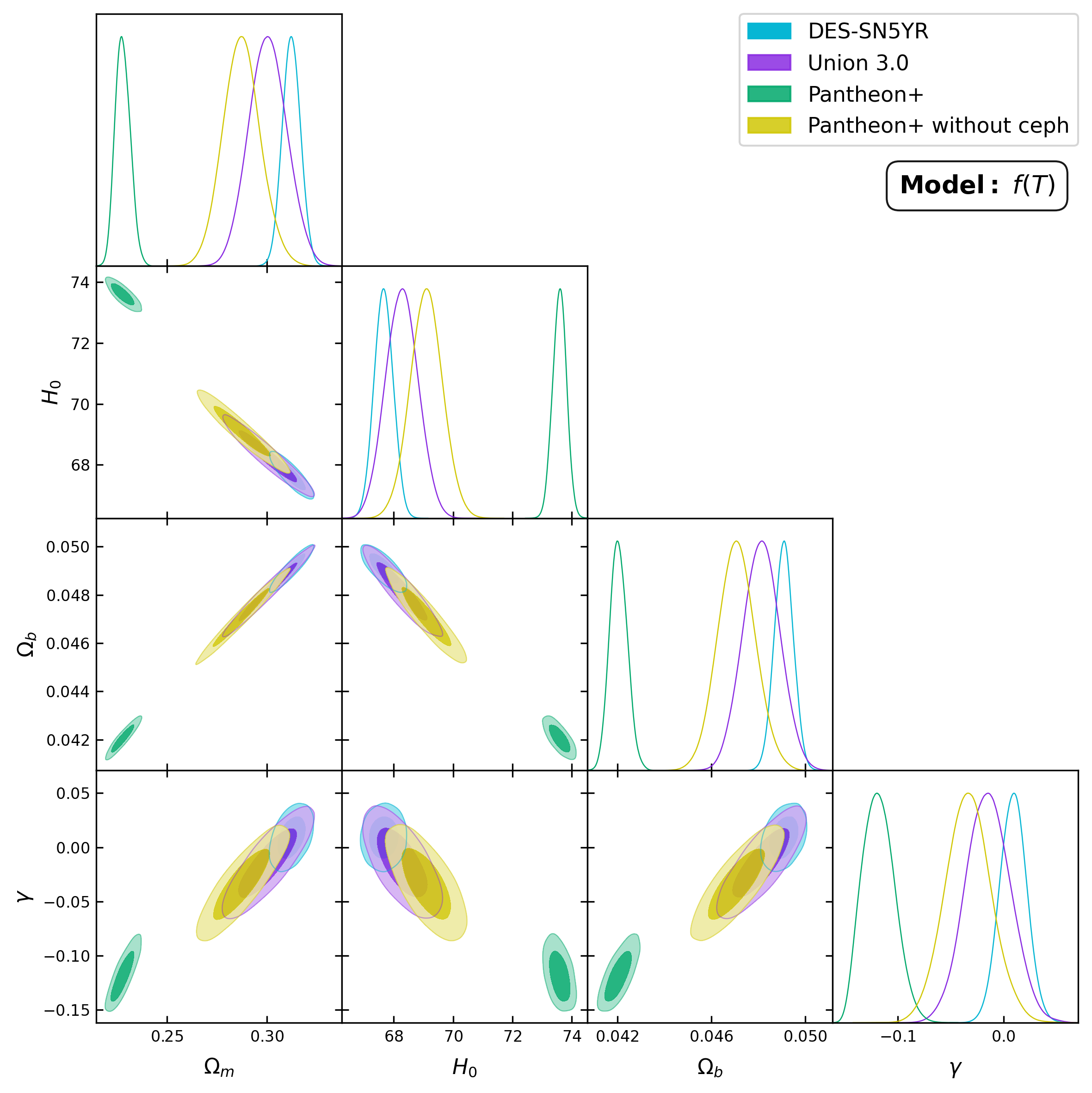}
    \caption{Marginalized posterior distributions and confidence contours for the parameters $\Omega_m$, $H_0$, $\Omega_b$ and $\gamma$ of the $f(T)$ model, obtained from CC, DESI DR2 BAO and Planck 2018 distance priors combined with each supernova sample: DES-SN5YR, Pantheon+SH0ES (with Cepheid calibration), Pantheon+ without Cepheid calibration, and Union~3.0. The diagonal panels show the one-dimensional posteriors and the off-diagonal panels show the $68\%$ and $95\%$ confidence regions. The Pantheon+SH0ES contours are displaced toward higher $H_0$, lower $\Omega_m$ and $\gamma\neq0$ along the degeneracy direction among these parameters, whereas the Pantheon+ analysis without Cepheids is consistent with $\gamma=0$ and lies close to the DES-SN5YR and Union~3.0 contours, indicating that the displacement originates from the Cepheid calibration.}
    \label{fig:placeholder}
\end{figure}

\subsection{Effective fluid evolution}
\label{subsec:effective_fluid}

Figures~\ref{fig:des_fluid},~\ref{fig:union_fluid} and~\ref{fig:pantheon_fluid} display the redshift evolution of the effective dark-energy fluid associated with the $f(T)$ correction, reconstructed for $0\leq z\leq 2$ using the best-fit parameters of each dataset listed in Table~\ref{tab:models_comparison}. In each figure, the left panel shows the effective energy density $\rho_{\rm eff}(z)$, the middle panel the effective pressure $p_{\rm eff}(z)$, and the right panel the effective equation-of-state parameter $w_{\rm eff}(z)=p_{\rm eff}/\rho_{\rm eff}$. Densities and pressures are expressed in units of the present-day critical density. Several features are common to the three cases. First, $\rho_{\rm eff}$ decreases monotonically with redshift, i.e., the effective component becomes progressively less important in the past, as expected for a late-time accelerating contribution. Second, its present value is fixed by the closure relation, $\rho_{\rm eff}(0)\simeq 1 - \Omega_{0r} -\Omega_{0m}$, so the differences at $z=0$ among the figures directly reflect the distinct matter densities preferred by each supernova sample, namely $\Omega_m=0.3125$ (DES-SN5YR), $0.2277$ (Pantheon+SH0ES) and $0.3001$ (Union~3.0). Third, the pressure is negative throughout, with $p_{\rm eff}(0)\approx-\rho_{\rm eff}(0)$, and $|p_{\rm eff}|$ declines toward zero at high redshift, so that the fluid mimics a cosmological constant today and fades out in the matter-dominated epoch.

\begin{center}
  \includegraphics[width=0.9\linewidth]{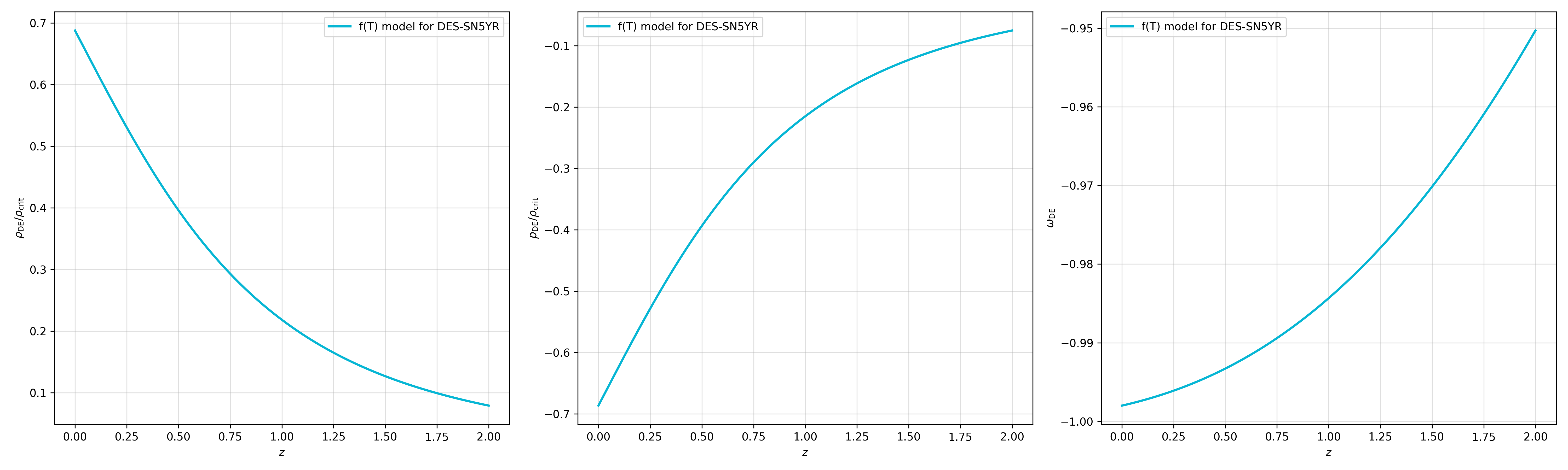}
  \captionof{figure}{Evolution of the effective geometric fluid of the $f(T)$ model as a function of redshift, using the best-fit parameters obtained with the DES-SN5YR sample. Left: effective energy density $\rho_{\rm eff}(z)$. Center: effective pressure $p_{\rm eff}(z)$. Right: effective equation of state $w_{\rm eff}(z)=p_{\rm eff}/\rho_{\rm eff}$, which starts at $-1$ and remains in the quintessence-like regime ($w_{\rm eff}>-1$) up to $z=2$.}
  \label{fig:des_fluid}
\end{center}

For DES-SN5YR (Fig.~\ref{fig:des_fluid}), the density decreases from $\rho_{\rm eff}(0)\approx0.69$ to $\approx0.08$ at $z=2$, and the pressure rises from $\approx-0.69$ to $\approx-0.1$. The equation of state starts at $w_{\rm eff}(0)=-1$ and grows smoothly and monotonically to $w_{\rm eff}\approx-0.95$ at $z=2$, remaining in the quintessence-like domain ($w_{\rm eff}>-1$) at all redshifts. This is a very mild departure from the cosmological constant, consistent with the best-fit value $\gamma=0.0092^{+0.0127}_{-0.0129}$, which is compatible with zero at the $1\sigma$ level. Accordingly, the improvement in the fit with respect to $\Lambda$CDM is marginal ($\chi^2_{\min}=1679.624$ against $1680.155$, i.e., $\Delta\chi^2\approx0.5$ for one additional parameter), and the inferred $H_0=67.65^{+0.32}_{-0.32}\,\rm km\,s^{-1}\,Mpc^{-1}$ is practically identical to the $\Lambda$CDM value. For this sample, the $f(T)$ fluid is therefore observationally almost indistinguishable from a cosmological constant.

\begin{center}
  \includegraphics[width=0.9\linewidth]{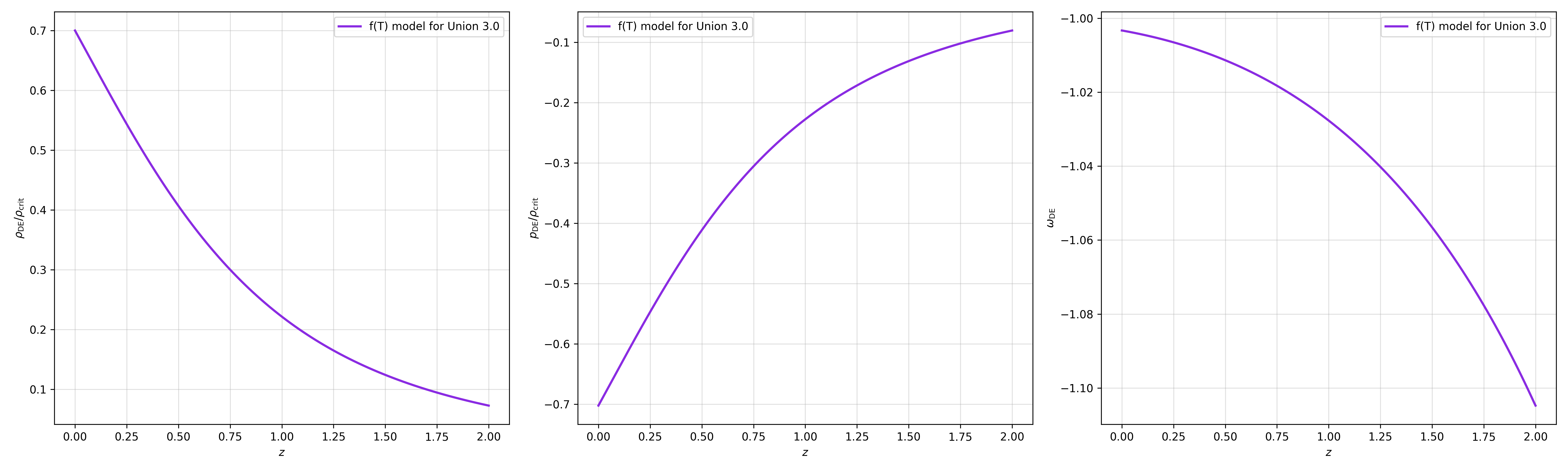}
  \captionof{figure}{Same as Fig.~\ref{fig:des_fluid}, but for the best-fit parameters obtained with the Union~3.0 sample. Left: effective energy density $\rho_{\rm eff}(z)$. Center: effective pressure $p_{\rm eff}(z)$. Right: effective equation of state $w_{\rm eff}(z)$, which decreases from $-1$ to $\approx-1.10$ at $z=2$, lying in the phantom regime ($w_{\rm eff}<-1$) for $z>0$.}
  \label{fig:union_fluid}
\end{center}

The Union~3.0 sample (Fig.~\ref{fig:union_fluid}) behaves similarly in the density and pressure panels, with $\rho_{\rm eff}$ decreasing from $\approx0.70$ to $\approx0.07$ and $p_{\rm eff}$ rising from $\approx-0.70$ to $\approx-0.07$ over the same interval. The main difference lies in the sign of the deviation of the equation of state: $w_{\rm eff}$ starts at $-1$ and decreases smoothly to $\approx-1.10$ at $z=2$, thus lying in the phantom regime ($w_{\rm eff}<-1$) at $z>0$. The magnitude of the deviation is small and is not statistically significant, given $\gamma=-0.0146^{+0.0215}_{-0.0210}$, which is again consistent with zero. The $\chi^2_{\min}$ values ($59.955$ for $f(T)$ and $60.56$ for $\Lambda$CDM) confirm that the data do not demand a dynamical component, while the larger uncertainties on $\Omega_m$ and $H_0$ compared with DES-SN5YR reflect the smaller size and lower statistical weight of this compilation.

\begin{center}
  \includegraphics[width=0.9\linewidth]{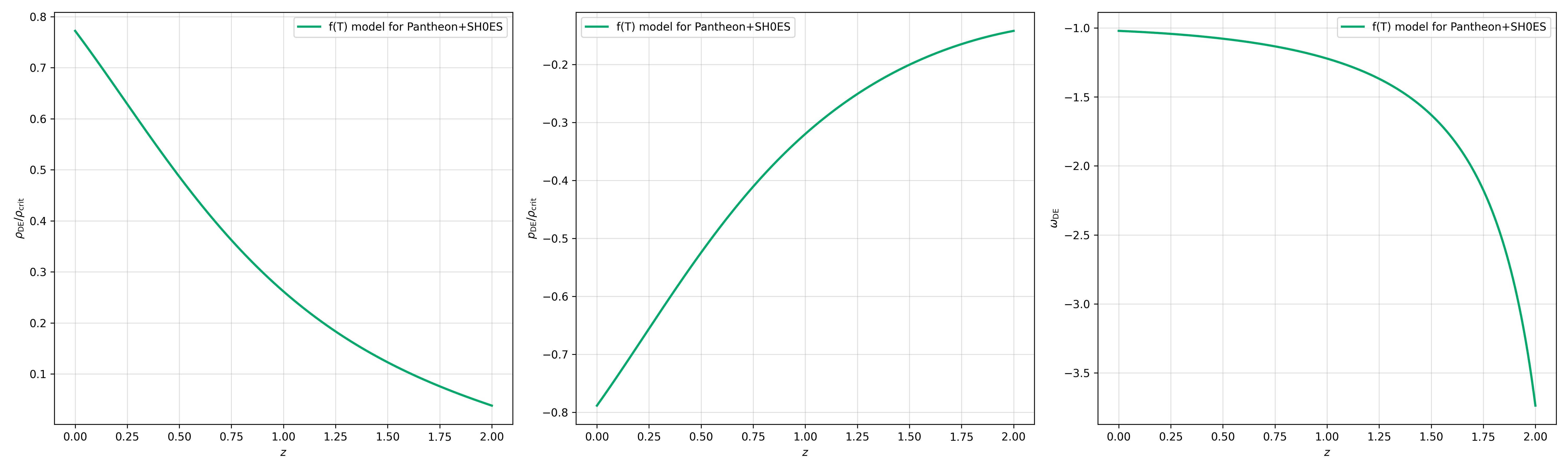}
  \captionof{figure}{Same as Fig.~\ref{fig:des_fluid}, but for the best-fit parameters obtained with the Pantheon+SH0ES sample. Left: effective energy density $\rho_{\rm eff}(z)$. Center: effective pressure $p_{\rm eff}(z)$. Right: effective equation of state $w_{\rm eff}(z)$, which remains close to $-1$ for $z\lesssim1$ and then drops steeply to $\approx-1.6$ at $z=2$, showing the most pronounced phantom behavior among the three samples.}
  \label{fig:pantheon_fluid}
\end{center}

The Pantheon+SH0ES sample (Fig.~\ref{fig:pantheon_fluid}) yields the most pronounced dynamical behavior. The effective density starts at the largest value among the three cases, $\rho_{\rm eff}(0)\approx0.77$, and drops faster, reaching $\approx0.03$ at $z=2$, while the pressure rises from $\approx-0.77$ to $\approx-0.05$. The equation of state remains very close to $-1$ for $z\lesssim1$ and then drops steeply, reaching $w_{\rm eff}\approx-1.6$ at $z=2$. This marked phantom behavior is driven by the clearly non-zero value $\gamma=-0.1190^{+0.0158}_{-0.0152}$, which deviates from the $\Lambda$CDM limit $\gamma=0$ by roughly $7\sigma$. It is accompanied by a substantial improvement in the fit, $\chi^2_{\min}=634.22$ for $f(T)$ against $683.55$ for $\Lambda$CDM, and by a high Hubble constant, $H_0=73.59^{+0.24}_{-0.24}\,\rm km\,s^{-1}\,Mpc^{-1}$, in line with the local distance-ladder determination. We stress that $w_{\rm eff}<-1$ does not signal a pathology of the underlying theory, the effective fluid is a geometrical construct arising from the torsion corrections to the Friedmann equations, not a fundamental matter field, so the null energy condition need not apply to it. The phantom behavior deserves further comment. In $f(T)$ gravity, the effective equation of state may cross $w=-1$ without introducing ghosts, since the effective fluid is a geometrical construct and the null energy condition does not apply to it. The phenomenon, however, is not free of consequences. For Pantheon+SH0ES, $w_{\rm eff}$ reaches $\simeq-1.6$ at $z=2$, a value well outside the range typically allowed by independent data. Moreover, $\rho_{\rm eff}$ decreases with redshift in all three cases, so this extreme equation of state occurs where the effective component contributes only $\sim3\%$ of the critical density; its impact on the observables is thus small and poorly constrained by the data. We regard the result at high $z$ as an extrapolation of the best-fit model rather than a measurement, and note that it is also the regime where the linear expansion in $\gamma$ is least reliable.

Taken together, the three figures show that the phenomenology of the effective fluid is controlled almost entirely by the parameter $\gamma$. For DES-SN5YR and Union~3.0, where $\gamma$ is consistent with zero, the reconstructed fluid stays close to a cosmological constant, with $|w_{\rm eff}+1|\lesssim0.1$ over the whole range, and the $f(T)$ and $\Lambda$CDM models are statistically equivalent. For Pantheon+SH0ES, in contrast, the data favor a significant evolution of the equation of state, with a smooth crossing into the phantom regime at intermediate redshifts. The fact that the sign and amplitude of the deviation differ among the samples, quintessence-like for DES-SN5YR and phantom for Union~3.0 and Pantheon+SH0ES, indicates that the constraints on the modified-gravity sector depend on the supernova compilation, on its calibration (in particular the inclusion of SH0ES Cepheid information, which pushes $H_0$ upward) and on the interplay with the remaining probes (CC, BAO and CMB) in the joint analysis. We therefore regard the Pantheon+SH0ES result as a hint of dynamical dark energy within the $f(T)$ framework that requires confirmation with independent datasets, while the DES-SN5YR and Union~3.0 results show that the model reduces consistently to $\Lambda$CDM when $\gamma\to0$.

\subsection{Dataset dependence of the $\gamma$ constraint}

The constraints on $\gamma$ differ markedly among the supernova samples: $\gamma$ is consistent with zero for DES-SN5YR and Union~3.0, but deviates from it at $\sim 7\sigma$ for Pantheon+SH0ES (Table~\ref{tab:models_comparison}). Two features of the analysis can account for this behaviour. First, Pantheon+SH0ES includes the Cepheid calibration of the SN~Ia absolute magnitude, which drives $H_0$ upward and $\Omega_m$ downward in the $\Lambda$CDM fit ($H_0=72.78$ km s$^{-1}$ Mpc$^{-1}$, $\Omega_m=0.2533$), in tension with the CMB and BAO constraints. Second, the contours in Fig.~\ref{fig:corner_mod} show a strong degeneracy among $\Omega_m$, $H_0$ and $\gamma$, so that a positive $\gamma$ can compensate a lower $\Omega_m$ and a higher $H_0$ while keeping the CMB and BAO observables approximately fixed. In this sense, $\gamma$ acts as an additional degree of freedom that absorbs the tension between the calibrated SN~Ia data and the early-universe probes.

To test this interpretation, we repeated the Pantheon+ analysis without the Cepheid calibration, i.e., removing the SNe in Cepheid host galaxies and using only SNe in the Hubble flow, selected by $z_{\rm HD}>0.023$ ($N_{\rm SN}=1371$ events), in combination with CC, DESI DR2 BAO and Planck 2018 distance priors, exactly as in the other analyses. Without the Cepheid distances the supernovae no longer fix the absolute scale, so the SN absolute-magnitude offset $M$ is treated as a free nuisance parameter and analytically marginalized over with a flat prior, using the full statistical$+$systematic covariance matrix restricted to the selected sample. In this setup the SNe constrain only the shape of the expansion history, and the absolute scale, hence $H_0$, is set by the combination of CC, BAO and CMB. We obtain $\Omega_m=0.2874^{+0.0096}_{-0.0089}$, $H_0 = 69.09^{+0.54}_{-0.55}$ km s$^{-1}$ Mpc$^{-1}$, $\Omega_b=0.0471\pm0.0008$ and $\gamma = -0.0332^{+0.0217}_{-0.0209}$. The value $\gamma=0$ lies within $\sim1.6\sigma$ of the posterior, so there is no significant preference for $\gamma\neq0$, and $\gamma$ is compatible within $1\sigma$ with the values obtained for DES-SN5YR and Union~3.0. The parameters are also very close to those of the CMB+BAO+CC analysis without supernovae (Table~\ref{tab:models_probes}), which indicates that the uncalibrated Hubble-flow SNe are consistent with the early-universe constraints and do not push the fit toward the high $H_0$ and low $\Omega_m$ of Pantheon+SH0ES. This supports the conclusion that the preference for $\gamma\neq0$ in Pantheon+SH0ES originates from the Cepheid calibration and not from the expansion history itself. We therefore interpret the Pantheon+SH0ES result as a symptom of dataset tension rather than as evidence for torsion-induced dynamics.

\section{Statistical model comparison}
\label{sec:aicbic}

Since the $f(T)$ model has one additional free parameter $\gamma$ relative to $\Lambda$CDM, a lower $\chi^2_{\min}$ is expected by construction, and a fair comparison must penalize model complexity. We therefore adopt the Akaike and Bayesian information criteria \cite{Akaike1974,Schwarz1978},
\begin{equation}
    \mathrm{AIC} = \chi^2_{\min} + 2k, \qquad \mathrm{BIC} = \chi^2_{\min} + k\ln N,
\end{equation}
where $k$ is the number of free parameters ($k=3$ for $\Lambda$CDM and $k=4$ for $f(T)$) and $N$ is the number of data points. We report $\Delta\mathrm{AIC}$ and $\Delta\mathrm{BIC}$ relative to the model with the lowest value in each dataset combination, so that positive values indicate a preference for the other model. Following the usual interpretation scales \cite{Burnham2002,Kass1995}, $|\Delta| \lesssim 2$ indicates no meaningful preference, $2 \lesssim |\Delta| \lesssim 6$ weak-to-positive evidence, $6 \lesssim |\Delta| \lesssim 10$ strong evidence, and $|\Delta| \gtrsim 10$ very strong evidence.
\begin{table}[H]
\centering
\renewcommand{\arraystretch}{1.5}
\setlength{\tabcolsep}{5pt}
\small
\caption{Information criteria for $\Lambda$CDM and $f(T)$ with CC, DESI DR2 and Planck 2018 CMB data combined with: (I) DES-SN5YR, (II) Pantheon+SH0ES, (III) Union 3.0. $\Delta$ values are relative to the lowest AIC/BIC in each attempt.}
\label{tab:aicbic_sne}
\begin{tabular}{c c c c c c c c c}
\hline\hline
Attempt & Model & $N$ & $k$ & $\chi^2_{\min}$ & AIC & $\Delta$AIC & BIC & $\Delta$BIC \\
\hline
I   & $\Lambda$CDM & 1868 & 3 & 1680.155 & 1686.155 & 0.000   & 1702.753 & 0.000 \\
    & $f(T)$       & 1868 & 4 & 1679.631 & 1687.631 & +1.476  & 1709.761 & +7.008 \\
\hline
II  & $\Lambda$CDM & 1744 & 3 & 683.551  & 689.551  & +47.340 & 705.943  & +41.877 \\
    & $f(T)$       & 1744 & 4 & 634.211  & 642.211  & 0.000   & 664.066  & 0.000 \\
\hline
III & $\Lambda$CDM & 65   & 3 & 60.560   & 66.560   & 0.000   & 73.083   & 0.000 \\
    & $f(T)$       & 65   & 4 & 59.944   & 67.944   & +1.384  & 76.642   & +3.559 \\
\hline\hline
\end{tabular}
\end{table}

Table~\ref{tab:aicbic_sne} shows the comparison for the three supernova samples. For DES-SN5YR (I), the improvement in $\chi^2_{\min}$ is only $0.524$, which does not compensate for the extra parameter: $\Delta\mathrm{AIC}=+1.476$ and $\Delta\mathrm{BIC}=+7.008$ in favor of $\Lambda$CDM, which is statistically preferred, strongly so by the BIC. The same occurs for Union~3.0 (III), with $\Delta\mathrm{AIC}=+1.384$ and $\Delta\mathrm{BIC}=+3.559$, meaning the two models are indistinguishable by the AIC, with a mild preference for $\Lambda$CDM by the BIC. In both cases $\gamma$ is consistent with zero within $1\sigma$ (Table~\ref{tab:models_comparison}), so the data show no evidence for deviations from $\Lambda$CDM. In contrast, Pantheon+SH0ES (II) strongly favors $f(T)$, with $\Delta\mathrm{AIC}=+47.340$ and $\Delta\mathrm{BIC}=+41.877$ for $\Lambda$CDM relative to $f(T)$, both well beyond the very strong threshold. This result must be interpreted with caution: for this sample, the $\Lambda$CDM fit is driven toward a high $H_0$ ($72.78$ km s$^{-1}$ Mpc$^{-1}$) and a low $\Omega_m$ by the SH0ES Cepheid calibration, which is in tension with the CMB and BAO constraints. The large improvement in $\chi^2_{\min}$ ($\Delta\chi^2 = 49.34$) therefore likely reflects the ability of the extra parameter $\gamma$ to accommodate this tension, rather than a genuine detection of a torsion-induced effect.

\begin{table}[H]
\centering
\renewcommand{\arraystretch}{1.5}
\setlength{\tabcolsep}{5pt}
\small
\caption{Information criteria for the $\Lambda$CDM and $f(T)$ models using Planck 2018 CMB data alone and combined progressively with DESI DR2 BAO and Cosmic Chronometers (CC), without supernova data. $N$ is the number of data points and $k$ the number of free parameters. $\Delta$AIC and $\Delta$BIC are relative to the lowest value in each dataset combination. For the CMB-only case, $N<k$ for $f(T)$, so the criteria are not statistically meaningful and are shown only for completeness.}
\label{tab:aicbic_probes}
\begin{tabular}{l l c c c c c c c}
\hline\hline
Dataset & Model & $N$ & $k$ & $\chi^2_{\min}$ & AIC & $\Delta$AIC & BIC & $\Delta$BIC \\
\hline
CMB & $f(T)$ & 3 & 4 & 0.01 & 8.01 & 2.01 & 4.40 & 1.10 \\
 & $\Lambda$CDM & 3 & 3 & 0.004 & 6.004 & 0.00 & 3.30 & 0.00 \\
\hline
CMB + DESI DR2 & $f(T)$ & 12 & 4 & 12.86 & 20.86 & 0.00 & 22.80 & 0.00 \\
 & $\Lambda$CDM & 12 & 3 & 17.193 & 23.193 & 2.33 & 24.65 & 1.85 \\
\hline
CMB + DESI DR2 + CC & $f(T)$ & 43 & 4 & 28.264 & 36.264 & 0.00 & 43.31 & 0.10 \\
 & $\Lambda$CDM & 43 & 3 & 31.931 & 37.931 & 1.67 & 43.21 & 0.00 \\
\hline\hline
\end{tabular}
\end{table}

Table~\ref{tab:aicbic_probes} presents the comparison without supernovae. For the CMB alone, the number of data points ($N=3$) is smaller than the number of free parameters of the $f(T)$ model, so the information criteria are not meaningful and are shown only for completeness. For CMB + DESI DR2, the AIC gives $\Delta\mathrm{AIC}=2.33$ in favor of $f(T)$ and the BIC $\Delta\mathrm{BIC}=1.85$, both in the weak-evidence regime. Adding CC data reduces the preference to $\Delta\mathrm{AIC}=1.67$, while the BIC becomes statistically equivalent ($\Delta\mathrm{BIC}=0.10$, now slightly favoring $\Lambda$CDM). Overall, apart from the Pantheon+SH0ES case, no combination of data shows significant evidence for $f(T)$ over $\Lambda$CDM, and the results are consistent with $f(T)$ reproducing a $\Lambda$-like behavior with a competitive fit, but without a statistically significant improvement.

\subsection{Limitations of the analysis}

Some simplifications in our treatment should be kept in mind when interpreting the results. (i) For the BAO data, we treat each measurement as independent, ignoring the correlation $r_{M,H}$ listed in Table~1 between $D_M/r_d$ and $D_H/r_d$. Since these correlations are negative ($r_{M,H}\sim-0.4$ to $-0.5$), neglecting them may slightly affect the size of the contours; we expect the central values to be less affected, but a full treatment with the DESI covariance matrix is desirable. (ii) The CC likelihood does not include the systematic covariance associated with the stellar population models. (iii) The CMB enters only through compressed distance priors ($R$, $\ell_A$, $\omega_b$), which do not capture information on perturbations; consequently, the CMB-only fit has no residual degrees of freedom (Table~4). (iv) The baryon density $\Omega_b$ is varied with a prior from Planck on $\omega_b$, which affects $r_d$ and the BAO constraints. (v) The supernova analyses use the standardized distance moduli provided in each data release, together with the full statistical$+$systematic covariance matrix, so that the light-curve standardization is not refitted simultaneously with the cosmological parameters. For Pantheon+ without Cepheid calibration, and for the other SN samples, the only nuisance parameter is the constant offset $M$ in the distance modulus, which is analytically marginalized over with a flat prior; for Pantheon+SH0ES with Cepheid calibration, the absolute scale is fixed by the Cepheid distances and no offset is marginalized. Since the standardization was performed by the collaborations assuming a fiducial cosmology, a residual dependence on that choice may slightly affect the uncertainties on $\gamma$ and $H_0$.

\section{Conclusions}
\label{sec6}

In this work we proposed a new $f(T)$ teleparallel model motivated by a quasi-constant effective energy density, $\rho_{\rm geo}=-\alpha\log_{2+\gamma}(2+\gamma T/T_0)$, which reduces exactly to a cosmological constant for $\gamma\to0$. To first order in $\gamma$, the model modifies the $\Lambda$CDM expansion history through a single extra parameter, $E^2(a,\gamma)=E^2(a,0)+\gamma\,(E^2(a,0)-1)(1-\Omega_{0m}-\Omega_{0r})/(2\ln 2)$. We confronted it with Cosmic Chronometers, three Type Ia supernova compilations (DES-SN5YR, Pantheon+SH0ES and Union~3.0), DESI DR2 BAO and Planck~2018 distance priors.

The main result is that a geometric, torsion-induced component can mimic a cosmological constant without one being introduced explicitly. For DES-SN5YR and Union~3.0, $\gamma$ is consistent with zero within $1\sigma$, and $\Omega_m$, $H_0$ and $\Omega_b$ agree with their $\Lambda$CDM values within the uncertainties. The reconstructed effective fluid stays close to $w_{\rm eff}=-1$ at low redshift. The information criteria confirm that the data do not demand the extra parameter: they favor $\Lambda$CDM for these two samples ($\Delta{\rm AIC}\simeq1.4$--$1.5$ and $\Delta{\rm BIC}\simeq3.6$--$7.0$), and in the analyses without supernovae the preference for $f(T)$ is marginal by the AIC ($\Delta{\rm AIC}\lesssim2.3$) and absent by the BIC ($\Delta{\rm BIC}\lesssim1.9$).

The only exception is Pantheon+SH0ES, which yields $|\gamma|\simeq0.12$, a $\sim7\sigma$ departure from $\gamma=0$, and $\Delta{\rm AIC}\approx47$ in favor of $f(T)$. We showed that this preference is a dataset effect rather than a detection of torsion-induced dynamics. When the Cepheid calibration is removed and only Hubble-flow SNe ($z_{\rm HD}>0.023$) are used, in combination with CC, BAO and CMB, we obtain $\gamma=-0.0332^{+0.0217}_{-0.0209}$, compatible with zero at $\sim1.6\sigma$, together with $\Omega_m=0.2874^{+0.0096}_{-0.0089}$ and $H_0=69.09^{+0.54}_{-0.55}$ km s$^{-1}$ Mpc$^{-1}$. In the calibrated sample, $\gamma$ acts as an additional degree of freedom that absorbs the tension between the Cepheid-calibrated SNe and the CMB+BAO constraints, through its degeneracy with $\Omega_m$ and $H_0$.

Regarding the Hubble tension, the model mitigates it but does not resolve it. The shifts in $H_0$ are of order $1\sigma$ and come with larger uncertainties, reducing the tension with the SH0ES value from $\sim4.3\sigma$ to $\sim3.3$--$3.4\sigma$ for CMB+BAO(+CC) and from $4.5\sigma$ to $4.0\sigma$ for Union~3.0, but it remains above $3\sigma$ in every case. The agreement obtained with Pantheon+SH0ES is built into the calibration of the data and is not an independent resolution. Likewise, the phantom-like values of $w_{\rm eff}$ reached at $z\sim2$ should be read as an extrapolation of the best fit: they occur where the effective component contributes only a few percent of the critical density, are not constrained by the data, and lie in the regime where the first-order expansion in $\gamma$ is least reliable.

Some simplifications of our analysis (Section~\ref{sec:aicbic}) limit the quantitative reach of these results, namely the neglect of the BAO correlations and of the CC systematic covariance, and the use of compressed CMB priors. Future work should include the full DESI covariance matrix, the CC systematics, and the full CMB power spectrum through an Einstein--Boltzmann solver, which requires the perturbation equations of the model. Including growth data (e.g.\ $f\sigma_8$) would also test whether the model is distinguishable from $\Lambda$CDM beyond the background level. Until then, our results show that this $f(T)$ model is a viable one-parameter extension of $\Lambda$CDM, statistically equivalent to it on current data, with $\gamma\to0$ consistently recovered once the calibration-driven tension is removed.

\section*{Acknowledgments}

The authors thank the Conselho Nacional de Desenvolvimento Científico e Tecnológico (CNPq), Brazil (Grant No. 141416/2023-8) (AOA), and JEGS (Grant No. 310560/2025-0).

\bibliographystyle{unsrtnat}
\bibliography{refs}

\end{document}